%% file: paper.tex
\documentclass[pdflatex,sn-mathphys]{sn-jnl}

\jyear{2021}%

\theoremstyle{thmstyleone}%
\theoremstyle{thmstyletwo}%

\theoremstyle{thmstylethree}%

\newcommand{\rev}[1]{{\color{black}#1}}

\usepackage{fontawesome}
\usepackage{amsmath,amssymb,amsfonts}
\usepackage{graphicx}
\usepackage{textcomp}
\usepackage{xcolor}
\usepackage{soul}
\usepackage{url}
\usepackage[utf8]{inputenc}
\usepackage{framed}
\usepackage{mdframed}

\begin{document}

\title[Requirements Evolution during Elicitation]{How Do Requirements  Evolve During Elicitation? An Empirical Study Combining Interviews and App Store Analysis}


\author*[1]{\fnm{Alessio} \sur{Ferrari}}\email{alessio.ferrari@isti.cnr.it}

\author*[2]{\fnm{Paola} \sur{Spoletini}}\email{pspoleti@kennesaw.edu}

\author[2]{\fnm{Sourav} \sur{Debnath}}\email{sdebnath@students.kennesaw.edu}

\affil*[1]{\orgdiv{Institute of Information Science and Technologies (ISTI)}, \orgname{National Research Council (CNR)}, \orgaddress{\street{Via G. Moruzzi 1}, \city{Pisa}, \postcode{56126}, \country{Italy}}}

\affil[2]{\orgdiv{Department of Software Engineering and Game Development}, \orgname{Kennesaw State University}, \orgaddress{\street{1100 South Marietta Pkwy}, \city{Marietta}, \postcode{30060}, \state{GA}, \country{USA}}}

\newpage


\abstract{
Requirements are elicited from the customer and other stakeholders through an iterative process of interviews, prototyping, and other interactive sessions. 
Then, requirements can be further extended, based on the analysis of the features of competing products available on the market. Understanding how this process takes place can help to identify the contribution of the different elicitation phases, thereby allowing requirements analysts to better distribute their resources. In this work, we empirically study in which way requirements get transformed from initial ideas into documented needs, and then evolve based on the inspiration coming from similar products. To this end, we select 30 subjects that act as requirements analysts, and we perform interview-based elicitation sessions with a fictional customer. After the sessions, the analysts produce a first set of requirements for the system. 
Then, they are required to search similar products in the app stores, and extend the requirements, inspired by the identified apps. 
The requirements documented at each step are evaluated, to assess to which extent and in which way the initial idea evolved throughout the process. 
Our results show that only between 30\% and 38\% of the requirements produced after the interviews include content that can be fully traced to initial customer's ideas. The rest of the content is dedicated to new requirements, and up to 21\% of it belongs to completely novel topics. Furthermore, up to 42\% of the requirements inspired by the app stores cover additional features compared to the ones identified after the interviews. The results empirically show that requirements are not elicited in strict sense, but actually co-created \rev{through interviews}, with analysts playing a crucial role in the process. In addition, we show  evidence that app store-inspired elicitation 
can be particularly beneficial to complete the requirements. 
}

\keywords{requirements engineering, requirements elicitation, interviews, app store, user stories, app store-inspired elicitation}



\maketitle

\input{sections/intro}

\input{sections/background}

\input{sections/method}

\input{sections/results}

\input{sections/discussion}

\input{sections/threats}
\input{sections/conclusion}

\backmatter




\bmhead{Acknowledgments}

This work was partially supported by National MIUR-PRIN 2020TL3X8X project T-LADIES (Typeful Language Adaptation for Dynamic, Interacting and Evolving Systems) and by the National Science Foundation under grant CCF-1718377.

\bibliography{bibliography}


\end{document}

%% file: sections/intro.tex
\section{Introduction}
Requirements are elicited from the customer and other stakeholders through an iterative process of interviews, prototyping, and other interactive sessions~\cite{zowghi2005requirements,dieste2010systematic,fernandez2017naming,PalomaresFQCLG21}. The iterations transform the initial ideas of the customer into more explicit needs, normally expressed in the form of a requirements document to be used for system specification. Then, the  analysis of competing products in the market~\cite{sarro2015feature,chen2019recommending,liu2019information,jiang2019recommending,al2019app}---typically retrieved from app stores~\cite{al2019app}---as well as the collection of feedback from users~\cite{maalej2016automatic,martin2016survey,jha2019mining,wang2019systematic}, can lead to further updates of the requirements, bug fixing and product enhancement.
Throughout the elicitation process, the initial customer's ideas can go through a radical transformation. Relevant needs may have remained unexpressed, others may have been discarded through early negotiation, and some requirements may be introduced to mimic successful features of existing products. Understanding how this process takes place can help to properly allocate resources for the elicitation activities, and also address possible communication problems typically occurring during the early elicitation phases~\cite{ferrari2017interview,salger2013requirements,bano2019teaching}. 

Previous literature in requirements engineering (RE) individually studied the different elicitation phases and supporting techniques.
Inquisitive elicitation strategies, such as interviews and focus groups, are recognised as extremely common in industrial practice~\cite{fernandez2017naming,PalomaresFQCLG21}. Some researchers studied the impact of domain knowledge~\cite{Hadar2014Role,NiknafsB17}, communication issues~\cite{ferrari2016ambiguity,bano2019teaching,ferrari2020sapeer} and other factors~\cite{coughlan2002effective,Davis2006,distanont2012engagement} on the success of these strategies. Concerning requirements elicitation through app store analysis, the survey by Al-Subaihin et al.~\cite{al2019app} highlighted that 51\% of mobile app developers frequently perform product maintenance based on user’s public feedback in the app stores, and 56\% elicit requirements by browsing similar apps. \rev{In the field of app store analysis}, a large set of work is dedicated to the development of automatic tools, often based on natural language processing (NLP), to support these tasks~\cite{maalej2016automatic,jiang2019recommending,wang2019systematic,zhao2021natural}. Despite the vast literature, however, it is unclear what is the actual contribution of the different elicitation activities to the final content of the requirements document. \rev{Furthermore, the recent survey by Dabrowski \textit{et al.}~\cite{dkabrowski2022analysing}, highlighted that---except for the study by Al-Subaihin et al.~\cite{al2019app}---there is limited evidence of the practical utility of app store analysis for RE and software engineering activities.}





In this work, we perform an exploratory study of descriptive nature to understand in which way requirements get transformed from initial ideas into documented needs, and then further evolve based on the analysis of existing similar products available on the market \rev{in general, and in the app store in the specific}. \rev{The study focuses on software solutions that can have a mobile version, or include a relevant mobile app-oriented component, so that comparison with products in the app store market can be possible.} 
The study consists of two phases. In Phase I, which we call \textit{interview-based elicitation}, we investigate the difference between customer ideas and documented requirements after a set of interview sessions. In Phase II, which we call \textit{app store-inspired elicitation}, we compare the  requirements produced as output of Phase I with additional ones created by analysts after taking inspiration from similar products identified in the app stores. 

To collect data for the study, we recruit 58 subjects that will act as requirements analysts. In the interview-based elicitation phase, the analysts perform a set of elicitation sessions with a fictional customer. The fictional customer is required to study a set of about 50 user stories for a system, which are regarded as the initial customer ideas for the experiment. Then, each analyst performs two requirements elicitation interview sessions with the customer, who is required to answer based on the user stories, and on novel ideas that can be triggered during the conversation. The sessions are separated by a period of 2 weeks, in which the analyst is working on the data collected in the first interview through notes, diagrams, or mockups.
After the elicitation sessions, each analyst documents the requirements into 50 to 60 user stories. 

Then, the app store-inspired elicitation phase takes place. The analysts are required to imagine that the initial idea needs to be evolved to reach a wider market, and thus additional requirements are to be added. To take inspiration for additional features, they are required to look into the app stores, e.g., Apple App Store and Google Play, and identify further requirements by analysing five similar products of their choice. After this analysis, they need to document the additional requirements with 20 user stories, and to explain why certain apps have been chosen as inspiration for the additional requirements.  

The produced user stories from a sample of 30 subjects are compared with the original ones by two researchers, to assess to which extent and in which way the initial requirements evolved throughout the interactive sessions. A comparison is also carried out between the user stories initially produced by the analysts, and those added after app store-inspired elicitation. \rev{In the comparison of user stories, we also consider differences in the user story roles, i.e., the actors explicitly mentioned in the user stories.} Our results quantitatively show that there is a substantial gap between initial ideas and documented requirements, and that searching for similar apps is also crucial to identify novel features and eventually enhance the product. 

More specifically, we provide the following main findings, in relation to \textit{interview-based elicitation}:  
\begin{itemize}
    \item Only between 30\% and 38\% of the user stories produced after interview-based elicitation include content that \rev{was already present in} the initial ones;
    \item Most of the user stories produced---specifically between 54\% and 63\%---are refinements of the initial ideas, while between 12\% and 20\% are related to completely novel ideas;
    \item Most of the \rev{user story} roles considered---between 58\% and 72\%---are the same as the ones initially planned, but between 18\% and 29\% are entirely novel roles;
    \item The relevance given to certain requirement categories is different between initial ideas and documented needs;
    \item The original \rev{user story}  roles are mostly preserved in the analysts' user stories, but the distribution of the analysts' stories among roles differs;
    \item Analysts introduce non-functional requirements, which were not initially considered by the customer, especially concerning \textit{security} and \textit{privacy}.
\end{itemize}

In addition, we provide the following findings in relation to \textit{app store-inspired elicitation}:

\begin{itemize}
    \item Between  28\%  and  42\%  of  the  user  stories  inspired  by the app stores cover entirely novel topics with respect to the ones written after the interviews;
    \item Between 8\% and 17\% novel roles are discovered;
    \item \rev{There is an increasing attention to the ``user'', i.e., the target subject who will be using the software, and \textit{usability} requirements, rather than to the initial ideas of the customer, i.e., the subject commissioning the initial development}. 
\end{itemize}

\rev{This study contributes to \textit{theory} in RE, as it empirically shows that: 1)~in interview-based elicitation, requirements are not elicited but \textit{co-created} by stakeholders and analysts; 2) interviews and app store analysis play
complementary roles in requirements definition; 3) app store analysis provides a relevant contribution in practice. This last point addresses the limited evidence about the practical utility of app store analysis observed by the survey of Dabrowski \textit{et al.}~\cite{dkabrowski2022analysing}.

The data of our study are shared in a replication package~\cite{replicationpkg2022} made available at:  \url{https://doi.org/10.5281/zenodo.6475039}}.

This paper is an extension of a previous conference contribution~\cite{DebnathS021}. The previous manuscript was focused solely on the evolution of requirements during the interview sessions. The current one, instead, includes also an analysis on the contribution of app store-inspired elicitation to the final requirements. 
More specifically, the current paper provides the following additional content: i) an extended description of the design of the study, including data collection and analysis for app store-inspired elicitation; ii) corresponding results and discussion related to app store-inspired elicitation; iii) an extended related work section; iv) additional statistical tests for the data of the previous contribution. The formatting and presentation of data, figures and original content has also been revised.


The remainder of the paper is organised as follows. Section~\ref{sect:background} summarizes the related work. Section~\ref{sect:method} describes the adopted methodology, Section~\ref{sec:rq1results} and~\ref{sec:rq2results} presents the results of our analysis and Section~\ref{sect:discussion} discusses their implications. Section~\ref{sect:threats} describes the threats to the validity of our results and how they have been mitigated and Section~\ref{sect:conclusion} concludes the paper.

%% file: sections/background.tex
\section{Related Work}
\label{sect:background}

Our work focuses on the evolution of requirements during  elicitation. In the following, we report related work on  requirements evolution in relation to early interview-based elicitation, and in relation to later phases, including app store-inspired elicitation. As the creativity of the analyst can influence the final requirements document, we also report related work on creativity in requirements engineering (RE). 
Finally, we discuss our contribution with respect to the literature in these research areas. 

\paragraph{Early Requirements Evolution}
Requirements evolution is a well-recognized phenomenon that can have critical effects on software  systems~\cite{harker1993evolution,crowne2002evolution,grubb2016evolution}. A few studies in the literature focus on requirements evolution during the early stages, towards the production of a first requirements document. Among them, Zowghi and Gervasi~\cite{ZOWGHI2003interplay} analyze how the initial incomplete knowledge about a system evolves and identify consistency, completeness, and correctness (the ``three C'') as the driving factors. The main idea behind this conclusion is that the goal of an analyst is to produce a \rev{consistent, complete} and consequently correct set of requirements. Thus, the analysts keep working using the collected knowledge and their expertise trying to get closer to the three Cs at every step of the process.
Grubb and Chechik~\cite{grubb2016evolution,grubb21Formal} focus on the early stage of requirements evolution, but they look at the modeling phase. In particular, they propose to use goal model analysis to help stakeholders to answer \textit{what if} questions to support the evolution of a system considering different scenarios, as well as the customer in understanding trade-offs among different decisions. In their approach, the authors augment goal models with the capability of explicitly modeling time to provide a more useful analysis for the stakeholders. Still within the field of model-driven RE, Ali \textit{et. al.}~\cite{Ali2011Requirements} identify in ``assumptions'' one of the main reasons behind requirements evolution.  Indeed, assumptions might be or can become inaccurate or incorrect. Once a problem with an assumption is identified, requirements need to evolve consequently. The authors develop a system to monitor the assumptions and evolve the model of the systems every time they are violated.  
Other authors have looked into the concept of \textit{pre-requirements}~\cite{hayes2008prereqir}, intended as information available prior to requirement specification----including system concepts, user expectations, the environment of the system----and their tracing with expressed needs. Studies in this field, and in particular Hayes \textit{et al.}~\cite{hayes2008prereqir} specifically focus on automatic tracing by means of cluster analysis.  

Another stream of works on early requirements elicitation is concerned with empirical studies on interviews and focus groups. These are the most common techniques used by companies, as shown by the NaPiRE survey~\cite{fernandez2017naming}, and by the recent study by Paolmares et al.~\cite{PalomaresFQCLG21}. Hadar \textit{et al.}~\cite{Hadar2014Role} show the double-sided impact of the domain knowledge of the analyst on the interview process. On the one hand domain knowledge can facilitate the creation of a shared understanding, but on the other hand it can lead to tunnel vision, with the analyst discarding relevant information. Niknas and Berry~\cite{NiknafsB17} show that domain ignorance can play a complementary role in requirements elicitation. By including a subject who is domain ignorant together with a software engineering expert in a requirements focus group, the process of idea generation appears to be more effective. Other studies on requirements elicitation interviews are focused on communication aspects. Among them, Ferrari \textit{et al.}~\cite{ferrari2016ambiguity} study the role of ambiguity, while Bano \textit{et al.}~\cite{bano2019teaching} present a list of typical communication mistakes committed by novel requirements analysts, which can have an impact on the resulting requirements document. 

\paragraph{Late Requirements Evolution}
The term ``requirements evolution'' often refers to the evolution of requirements once the system is deployed, and a wide body of work exists in the area. In particular, a set of studies consider evolution of requirements based on the analysis of similar products, considering in particular app stores. Fu \textit{et al.}~\cite{fu2013people} acknowledge that a market-wide analysis across the entire app store can allow developers to find undiscovered requirements. In this line of research, Sarro \textit{et al.}~\cite{sarro2015feature} empirically shows that features migrate through the app store, passing from one product to another, and suggest that identifying those features that have strong migratory tendency can lead to identify undiscovered requirements. Other works are concerned with the development of recommender systems, e.g.,~\cite{chen2019recommending,liu2019information,jiang2019recommending,liu2021supporting}. Among them, Chen \textit{et al.}~\cite{chen2019recommending} recommend software features for mobile apps, based on the comparison of the user interface (UI) of an existing app with the UIs of similar ones. Liu \textit{et al.}~\cite{liu2019information}, instead leverages textual data in terms of app feature descriptions, to support feature recommendation. Jiang \textit{et al.}~\cite{jiang2019recommending} extend descriptions with API names, to better inform the recommendation system. Finally, the recent work of Liu \textit{et al.}~\cite{liu2021supporting} combines UI information and textual descriptions of apps. 
Besides the field of app store analysis, the topic of requirements elicitation based on similar products has been also studied in software product line engineering. The majority of the works in the area focuses on the automatic identification of product features from existing documents---brochures or requirements---by means of natural language processing (NLP) techniques~\cite{ferrari2013mining,li2017reverse,davril2013feature}. A literature review on similarity-based analysis of software applications is presented by Auch \textit{et al.}~\cite{auch2020similarity}. 

Another well-studied factor of late requirements evolution is user feedback~\cite{carreno2013userfeedback,martin2016survey}, in the form of app reviews, Tweets or other media. Carre\~no and Winbladh~\cite{carreno2013userfeedback} created a system to automatically extract topics from user feedback and generate new requirements for future versions of the app. Feedback in the form of app reviews is analysed by a stream of works from Maalej and his team (e.g.,~\cite{maalej2016automatic,pagano2013user}).
Along the same line of research, Guzman \textit{et al.}~\cite{Guzman2017Bird} proposes to use the information mined from Twitter to guide the evolution of requirements. Additional similar approaches are discussed by Khan \textit{et al.}~\cite{Khan2019Crowd} and by Morales-Ramirez \textit{et al.}\cite{Morales-Ramirez19}. 
 
 

\paragraph{Creativity in RE}
Creativity plays an important role in many RE activities, including the requirements evolution process~\cite{Robertson2005reinvention, lemos2012mappingcreativity, Maiden2010recreative}. 
According to Sternberg, creativity can be described as \textit{the ability to produce work that is both novel and appropriate}~\cite{sternberg1999handbook}. Nguyen~\cite{NGUYEN2009framework} states that creativity can be attributed to five factors: product, process, domain, people, and context. 
To analyze the impact of creativity in discovering new requirements, Maiden \textit{et al.}~\cite{Maiden2005Integrating} performed a study consisting of a series of creative workshops to discover new requirements for an Air Traffic Management System. 
This work provides empirical evidence of the impact of creativity and creative processes in identifying new requirements. Inspired by these seminal  contributions on creativity and RE, the CreaRE workshop was established\footnote{\url{https://creare.iese.de}}, and it is currently at its 10th edition, indicating the interest of the community in the topic. Several techniques have been experimented in the literature; among them, the EPMcreate technique~\cite{Herrmann2018Creativity},
theoretical frameworks for understanding creativity in RE~\cite{NGUYEN2009framework}, platforms
to support collaboration for distributed
teams~\cite{Mahaux2013Collaborative}, toolboxes for selecting the appropriate creativity technique~\cite{Grube2008Selecting, Svensson2015selecting}, and the use of combination of goal modeling and creativity techniques ~\cite{Horkoff2015Creativity, Horkoff2016Creative}.

\subsection{Contribution}

With respect to the literature on early requirements evolution, our work is among the first ones that considers requirements elicitation performed with traditional interviews, which are extremely common in practice~\cite{Davis2006,fernandez2017naming, PalomaresFQCLG21}. The closest work to ours is the contribution of Hayes \textit{et al.}~\cite{hayes2008prereqir}, focusing on pre-requirements information. Their concept of pre-requirement is analogous to our notion of ``initial idea''. However, their goal is to aggregate and automatically cluster pre-requirements from multiple stakeholders, and support traceability. Instead, in our paper we want to evaluate how the initial ideas get transformed through the elicitation and documentation process. 

Compared to work on late requirements evolution, most of the works focuses on \textit{automatic} tools for app store mining, supporting feature extraction or app review analysis. Instead, in this paper we consider a manual approach for app store-inspired elicitation, and we measure the impact with respect to existing requirements. 
Our paper also differs from most of the existing works because we do not consider the evolution of requirements \textit{after} a product has been developed. Instead, we study the extension of existing requirements based on a market analysis performed before releasing the product. 

Compared to work on creativity our study also differs from the literature.  Instead of providing a novel technique to stimulate creativity, it gives quantitative evidence on the impact of early elicitation and documentation activities. In particular, it shows that (a) creativity takes place as a natural phenomenon without introducing specific triggering techniques; (b) a quantitative evaluation is possible, and can be used to compare different creativity techniques for requirements elicitation.

%% file: sections/method.tex
\section{Research Design}
\label{sect:method}
\subsection{Research Questions}
The overarching objective of this research is to explore in which way requirements evolve from ideas to expressed needs. More specifically, we want to first understand what is the difference between initial ideas and user stories documented after initial interactions with a customer. Then, we want to understand and what is the difference between these user stories and those that come out after an analysis of similar products in the market. This way, we can understand how different strategies of requirements elicitation contribute to the definition of a product.  
Two main research questions, with associated sub-questions, are considered:

\begin{itemize}
\item RQ1: What is the difference between the initial customer ideas and the requirements documented by an analyst after customer-analyst interview sessions?

\begin{itemize}
   \item RQ1.1: \textit{How \rev{large} is the difference in terms of documented requirements and roles with respect to initial ideas?} With this question we want to quantitatively evaluate how different are the user stories with respect to initial customer ideas. This gives a numerical indication of how much is the contribution of the elicitation sessions to documented requirements. 
    \item RQ1.2: \textit{What is the relevance given to the different categories of requirements and roles with respect to initial ideas?} The question aims to understand whether there is a difference in terms of relevance given to categories or requirements with respect to initial customer ideas. \rev{With the term ``relevance'' we arguably intend the percentage of the stories dedicated to a certain category or role}\footnote{\rev{A single requirement could be practically more relevant than others. However, here we consider relevance of \textit{categories}---and not single requirements---in relation to other categories. Our usage of the term `relevance' assumes that more user stories in a category implies that the analyst considered that category more important than others.}}. In other terms, we want to understand whether the elicitation sessions \rev{gave more prominence} to certain aspects with respect to others, compared to initial ideas. 
    \item RQ1.3: \textit{What are the emerging categories and roles?} This question aims to understand whether there are typical categories of user stories and roles that were not originally present in the ideas of the customer, and therefore what is the actual contribution of the elicitation process in terms of content. 
\end{itemize}

\item RQ2: What is the additional contribution of requirements produced after an analysis of products available from the app stores?


\begin{itemize}
    \item RQ2.1: \textit{How \rev{large} is the difference in terms of covered requirements categories and roles with respect to the requirements documented after the interview sessions?} This question aims to quantitatively evaluate how much is the contribution given by the analysis of products from the app stores, and if the additional requirements introduced cover different categories with respect to the ones documented in the previous phase.  
    \item RQ2.2: \textit{What is the relevance given to the different categories of requirements and roles with respect to the requirements documented after the interview sessions?} This question investigates what are the specific requirements categories and roles considered in the additional requirements, and whether their distributions are different, when compared with the previously documented requirements. In other terms, we want to understand if the focus of the analysts shifted in this second phase. 
    \item RQ2.3: \textit{What are the additional categories and roles?} This question aims to identify whether there are categories and roles that are common across analysts, and that emerged in this specific phase. This allows us to understand what is the contribution of app store-inspired elicitation.
\end{itemize}

\end{itemize}


\begin{figure}[t]
\centering
\includegraphics[width=\columnwidth]{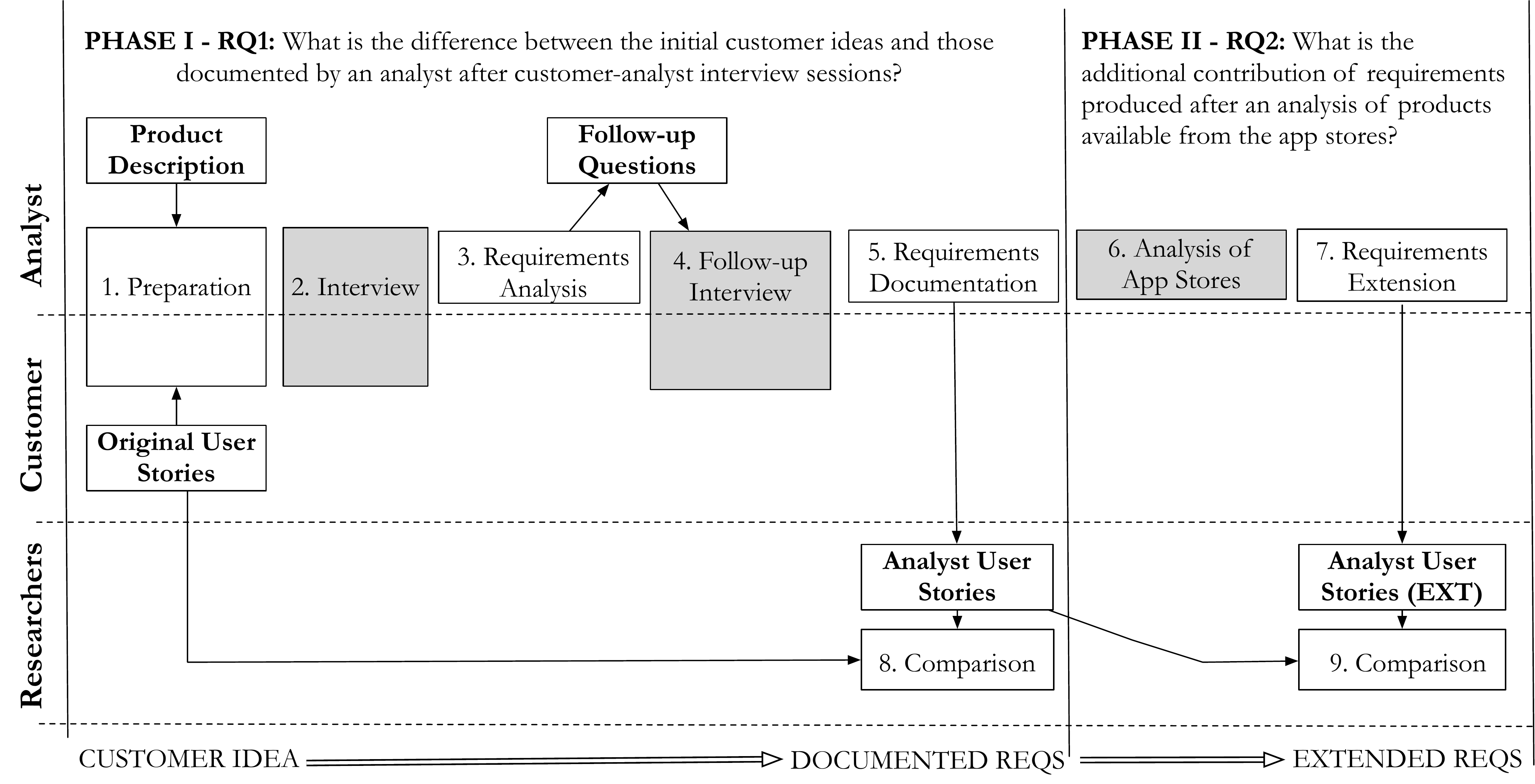}
\caption{Steps of the experiment.}
\label{fig:process}
\end{figure}

\subsection{Data Collection}
\label{sec:datacollection}

To answer the questions, we perform an experiment with one fictional customer (3rd author of the current paper), and a set of 58 different student analysts recruited from Kennesaw State University. \rev{In particular, the analysts were graduate students enrolled in the first or second semester of the MSc in Software Engineering and, at the time of the experiment, were all taking a course on Requirements Engineering. In that course, they have been introduced to elicitation techniques and user stories using very standard teaching material which also included examples from very common systems such as an event ticket store and a bookstore. During the first semesters of the program, students have usually taken, or are taking, an introductory course on software engineering and a course on project management. Elective specialized courses (e.g., User Interaction Engineering and courses on Privacy) are in general taken later in the program.}

The steps of the experiment (approved, together with the recruitment process, by the Institutional Review Board of Kennesaw State University) are described in the following. Please refer to Fig.~\ref{fig:process} for an overview of the steps. Overall, the study is composed of two phases. Phase I, which we call \textit{interview-based elicitation}, is associated to RQ1 and includes all the steps that go from initial customer ideas until documented requirements. Phase II, called \textit{app store-inspired elicitation}, is associated to RQ2, and includes the \rev{steps that lead} to an extension of the requirements based on the analysis of app stores. 

\paragraph{Phase I - Analyst and Customer Steps}

\textbf{1. Preparation} Analysts are given a brief description of a product to develop and are asked to prepare questions for a customer that they will have to interview to elicit the products’ requirements. The product is a system for the management of a summer camp. The brief description has an initial part that describes the company's current practice followed by a set of briefly described needs about (1) managing the information, registration, and activities of the participants, (2) giving the participants' guardians the opportunity to register and follow their children, (3) managing the employees’ performance and schedule, 
(4) communicating with parents and employees, and (5) managing facilities.\\
The fictional customer is required to study a set of about 50 user stories for the system, which are regarded as the initial customer ideas for the experiment. The initial user stories are taken from the dataset by Dalpiaz~\cite{dalpiaz2018}, file \texttt{g21.badcamp.txt}.
The use of the same customer for all the interviews is in line with similar experiments, such as \cite{bano2019teaching} and \cite{Pitts2004Stopping}.

\textbf{2. Interview} 
Each analyst performs a 15 minutes interview with the customer, possibly asking additional questions with respect to the ones that they prepared. The fictional customer answers their questions based on the set of user stories that describe the product, and that are not shown to the analysts. Overall, the customer is required to stick to the content of the user stories as much as possible. However, he is allowed to answer freely when he does not find a reasonable answer in the user story document, to keep the interview as realistic as possible, and capture novel ideas that emerge in the dialogue. The analysts are required to record their interviews, and take notes. 

\textbf{3. Requirements Analysis}
Based on the recording and their notes, the analysts have to: (a) perform an initial analysis of the requirements, and based on this analysis (b) produce additional questions for the customer to be asked in a follow-up interview. The initial analysis can be performed with the support of a graphical prototype, use cases, or written form. Analysts can adopt the method they find more suitable.

\textbf{4. Follow-up Interview}
Then, they perform a follow-up interview with the customer, which also lasts 15 minutes, to ask the additional questions prepared. During the interview, they can use the graphical prototype, the use cases, or any material produced as a support to ask questions to the customers. In practice, they can show the material to the customer and discuss based on the material. 

\textbf{5. Requirements Documentation}
After the second interview, they are required to write down from 50 to 60 user stories for the system. We constrain the number of user stories between 50 and 60 to be consistent with the number of user stories in the original set and, thus, to better compare and analyze the collected data. About 50 user stories are also the typical number in the dataset by Dalpiaz~\cite{dalpiaz2018}, which we deem representative of user story sets used for research purposes.

\paragraph{Phase II - Analyst and Customer Steps}

\textbf{6. Analysis of App Stores}
Each analyst starts from the user stories that they developed in the previous phase. They are now asked to produce an enhanced set of user stories, which take into account possible competing products available on the market, as commonly done by app developers~\cite{al2019app}. 
To this end, the analysts are required to perform an informal market analysis based on the app stores. In particular, they are required to:

\begin{itemize}
    \item Select at least 5 mobile apps from Google Play or the Apple App Store that are in some way related to the developed product (for example apps for summer camps, apps for trekking, or anything that they consider related).
    \item  Try out the apps to have an idea of their features when compared with their product. 
    \item Go through the app reviews to identify desired features, and additional requirements that may be appropriate also for their product. 
\end{itemize}

No automated tool, except for the default app store search engines, was provided for the market analysis task. \rev{The analysts were free to browse the app stores following their intuition, ability and preferred strategy.}

\textbf{7. Requirements Extension}
Based on the analysis of the app stores, the analysts are required to list the selected apps and their links, together with a brief description that 1. outlines the main features of the product, 2. explains in which way the product is related to the original one, and why they have chosen it. Furthermore, they are asked to add 20 user stories to the original list, based on the analysis of the app stores. 

\paragraph{Phase I - Researchers' Steps}

\textbf{8. Comparison with Initial Ideas}
Among the 58 participants, some did not consent to use their work for publication. In addition, as the analysis has been performed manually, to make it manageable and consistent, we reduced the number of analyzed analysts to 30 which is a number that still allowed us to obtain significant results. We randomly selected them and their work has been inspected by two researchers (2nd and 3rd author) to identify:

\begin{itemize}
\item User stories that express content that was already entirely present in the initial set of user stories (marked as ``existing'', \textbf{E}).
\item User stories that express content that is novel with respect to the initial set of user stories, but that belongs to one of the existing high-level categories of the initial set (marked as ``refinement'' \textbf{R}).
\item User stories that express content that is novel, and belonging to a novel category not initially present (marked as ``new'' \textbf{N}).
\item The name of novel categories of user stories introduced.
\item Recurrent themes in \textbf{R} and \textbf{N} stories.
\item Roles that were used also in the original stories (\textbf{E}$_\rho$).
 \item Roles that represented a refinement of roles used in the original stories (\textbf{R}$_\rho$).
 \item Roles that were novel and never considered in the original stories (\textbf{N}$_\rho$).
\end{itemize}

\begin{figure*}[t]
\centering
\includegraphics[width=\textwidth]{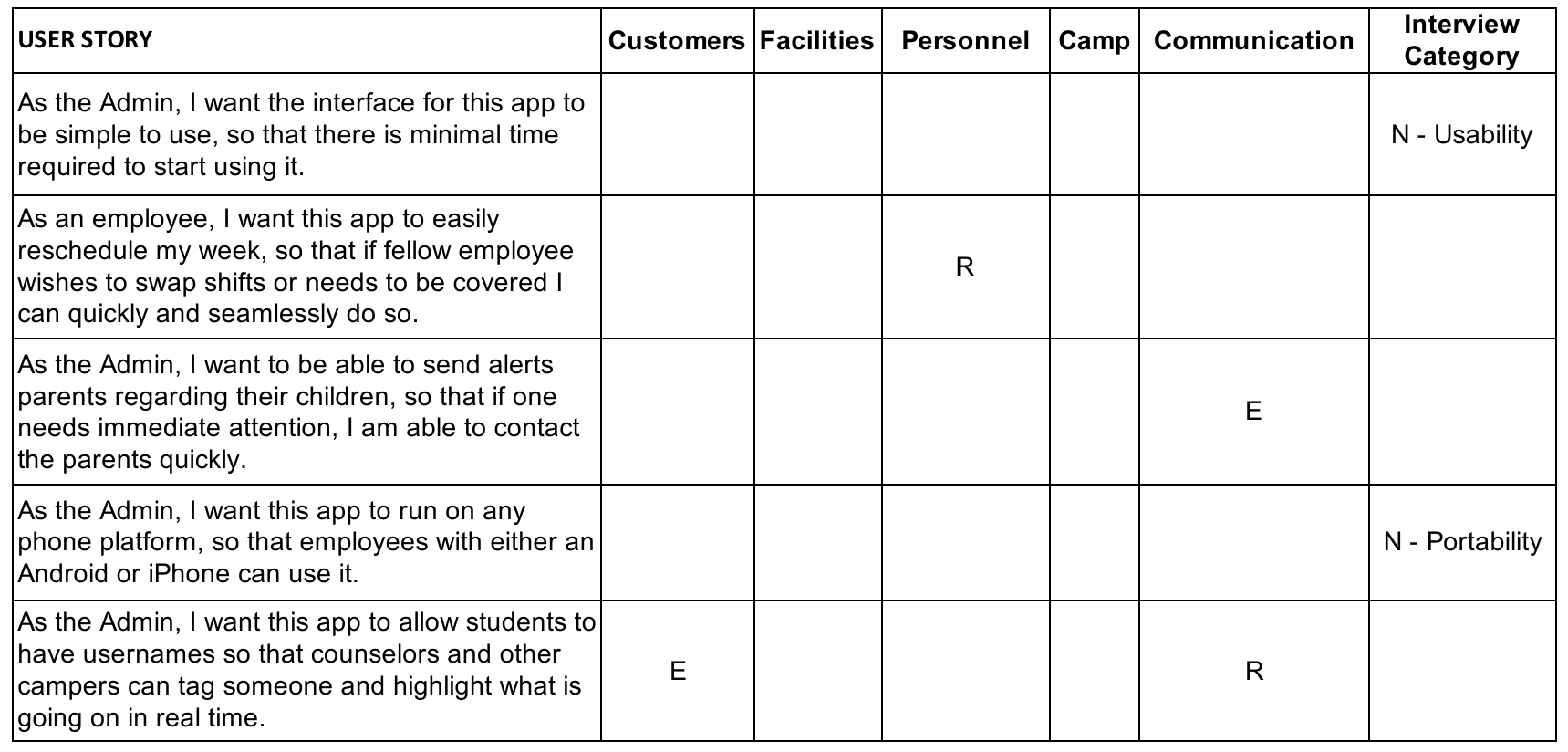}
\caption{Extract of a compiled spreadsheet.}
\label{fig:excerpt}
\end{figure*}

This process is carried out by means of a template spreadsheet that is used to annotate the user stories. Given a list of user stories produced by one of the analysts, a researcher went through the list, and marked each user story with E, R, or N. 
\rev{For E and R, the researcher was asked to mark the high-level original category to which the user story belonged. The high-level original categories were extracted according to the \textit{Validity Procedure}, reported later in this section, and are \textit{Customers}, \textit{Facilities}, \textit{Personnel}, \textit{Camp}, and \textit{Communication}.} 
Whenever a user story was marked with N, the researcher was asked to report the name of the new category identified. An excerpt of the data analysis for one of the user story documents is reported in Fig.~\ref{fig:excerpt}. \rev{It should be noticed that, to make the annotation task practically feasible, the evaluation was carried out considering high-level, domain-specific categories of user stories and not by linking individual user stories with original ones. This was considered hardly feasible, since single user stories could be traced to multiple ones, or to subsets of existing ones. This  fine-grained task would have led to major disagreements between annotators, thus hampering the validity of the results.}


Subsequently, the roles used in the stories were extracted and marked as E$_\rho$, R$_\rho$ or N$_\rho$. For the case of E$_\rho$ and R$_\rho$ roles, the researcher also indicated the corresponding role in the original story. For N$_\rho$, the role was added to the list of roles. 

\textit{Validity Procedure.} \rev{To produce the data analysis scheme presented in Fig.~\ref{fig:excerpt}, we needed to identify the \textit{original categories}---i.e., a categorisation for the original user stories---and \textit{interview categories}---i.e., the names of the novel categories emerging from the user stories produced after interviews. These latter are the possible values of cells under the column ``Interview Category'' in Fig.~\ref{fig:excerpt}. 
To identify these categories, a validity procedure was set-up, comprising the following steps. 
}

\begin{itemize}
\item \rev{\textbf{Identification of original categories:}} The initial set of user stories used as preparation material for the interviewee has been analyzed by the three authors independently \rev{to identify the high-level categories emerging from the original user stories, referred in the following as \textit{original categories}. These categories are:  \textit{Customers}, \textit{Facilities}, \textit{Personnel}, \textit{Camp}, and \textit{Communication}, and are defined in more detail in Sect.~\ref{sec:originalcategories}.}

\item \rev{\textbf{Validation of original categories:} To validate the original categories, each researcher independently used them to label a set of 107 user stories sampled from different analysts. If a user story did not belong to any of the original categories it was marked with ``N'' to indicate a new category, and a preliminary name was provided. After the labeling process, the researchers met in a 30 minutes meeting to reconcile the disagreements. For 9 of the user stories (about $8.5\%$ of the total), they either assigned different labels or they were not sure which label to assign and wanted to discuss the user story in the meeting.}

\item \rev{\textbf{Identification of interview categories:} The researchers met in a 1 hour meeting to agree on the names of the new categories emerging from the data, referred in the following as \textit{interview categories}. As one of the researchers also played the role of the customer in the interviews, he gave additional insights which helped to name the new categories. When further input was needed, the researchers also referred to the recordings of the interviews. After the meeting, the  spreadsheet model used for the comparison procedure was produced, and it is shown in Fig.~\ref{fig:excerpt}.}
\item \rev{\textbf{Consolidation of the schema:} Using the spreadsheet model the 2nd and 3rd authors independently labeled more than 100 user stories and then met to reconcile. In this activity, novel interview categories were also introduced. The disagreement on the category to assign to the user stories was minimal (less than $~4\%$ of the cases) and was mostly related to the name assigned by the researchers to new interview categories. The 1st author analyzed the spreadsheet model and approved the consolidated schema.}
\item \rev{\textbf{Application of the Schema:} The schema was applied to the whole set of user stories by the 2nd author to carry out the comparison activity. In this activity, the author could use the original categories, the interview categories identified in the previous steps, or add further interview categories. The user stories that raised doubts (about 2\%) were marked as ``to be discussed''.} 
\item \rev{\textbf{Final Reconciliation:} After the analysis a final meeting was held between 2nd and 3rd author, to consolidate the names of additional interview categories and resolve doubts. This meeting was broken over two days for a total of more than 6 hours.} 
\end{itemize}


\rev{
The analysis of the original roles did not require a similar effort, as the original user stories clearly identified three well-separated roles, namely \textit{\textit{Administrator}, \textit{Worker} and \textit{Parent}}. For the new roles emerging from interviews, the user stories also clearly identified role names, and these were discussed and confirmed in the final meetings.}

\paragraph{Phase II - Researchers' Steps}

\textbf{9. Comparison between Interview-based Elicitation and App Store-inspired Elicitation} The additional user stories of the 30 analysts already considered in Step 8 are evaluated by two researchers (2nd and 3rd author) to identify, for each analyst: 

\begin{itemize}
    \item The user stories that express content that belonged to novel categories not considered in the previous phase by the specific analyst (referred in the following as ``app-inspired'', \textbf{A}). 
    \item Roles that were novel with respect to those previously considered by the specific analyst (\textbf{A}$_\rho$).
\end{itemize}

\begin{figure*}[t]
\centering
\includegraphics[width=\textwidth]{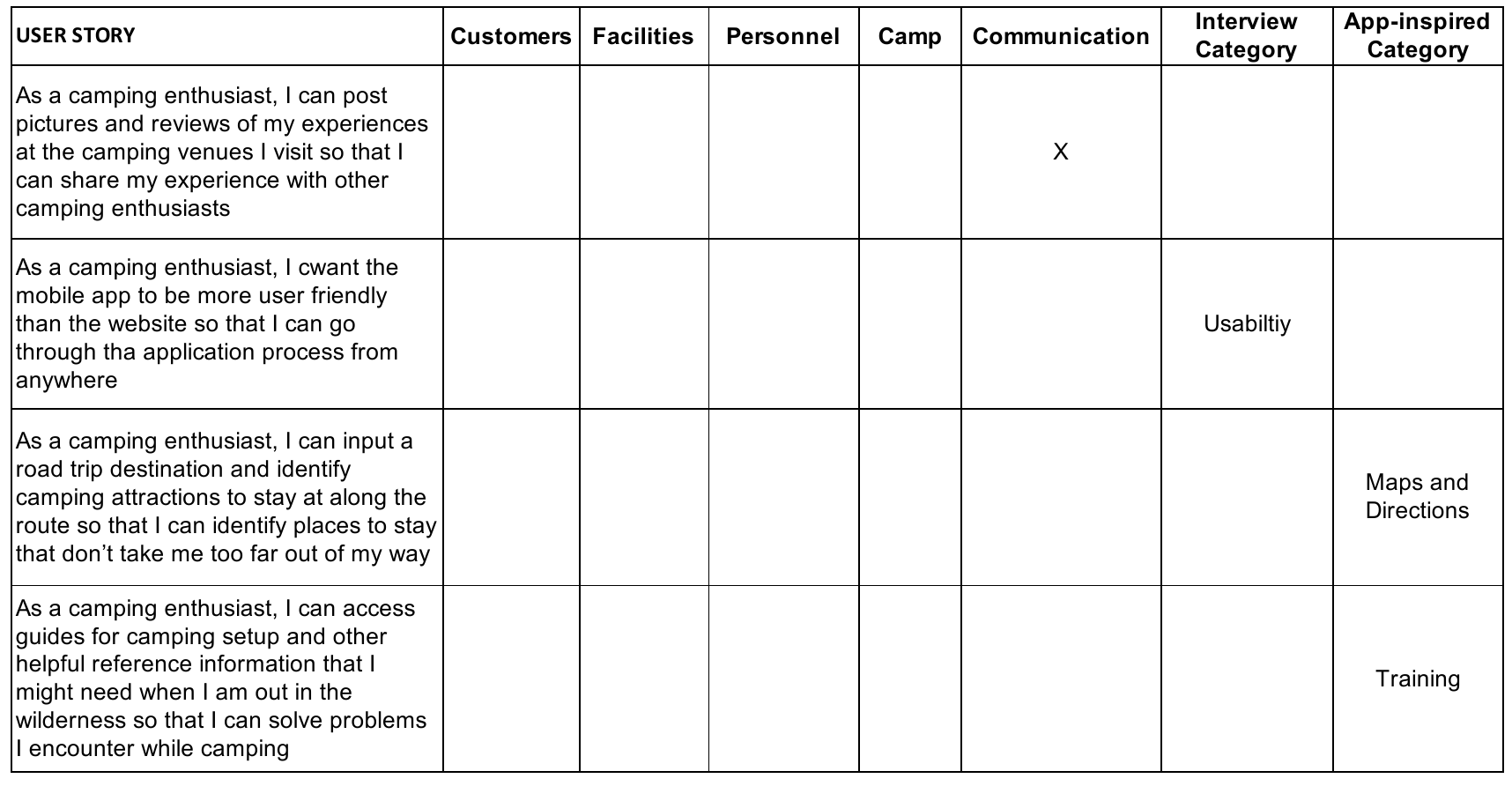}
\caption{Extract of a compiled spreadsheet for the second phase.}
\label{fig:ecerptappstores}
\end{figure*}

This process is supported by a template spreadsheet that is used to annotate the user stories. An example is reported in Fig.~\ref{fig:ecerptappstores}. 
For each list of user stories belonging to an analyst, a researcher went through the list, and assigned a category to each user story. The category could be selected among the five original ones, and also within the whole set of categories already identified in the user stories of \textit{all} the analysts in the previous phase, i.e., all the interview categories. If an appropriate category was not present in the list, the researcher could add it in an additional column (``App-inspired Category'', in Fig.~\ref{fig:ecerptappstores}). Since the researchers could chose among \textit{all} the categories---including original ones---they could also select categories that were not considered by the specific analyst in the initial phase. Therefore, to determine whether a user story could be marked as \textbf{A}, we automatically extracted the list of categories already present in the interview-based user stories of the specific analyst. A user story produced in this second phase was considered as \textbf{A} if its category did not appear in the extracted list. This way, even though a category was already considered by some other analyst, or was part of the original set, it would be counted as ``app-inspired'' for the specific analyst. 

After this analysis, the roles were extracted and marked as \textbf{A}$_\rho$, with the same rationale and approach used for the marking of user stories. 

\textit{Validity Procedure.} The alignment between researchers in terms of categories of user stories was already achieved in the previous step. In this step, we needed to ensure that the newly introduced categories, referred in the following as \textit{app-inspired categories}, were uniform and agreed upon. Therefore, the 2nd and 3rd authors independently annotated \rev{half of the complete list of user stories produced by the 30 analysts, around 300 user stories for each author}. Then, they cross-checked each other's work, to identify disagreements \rev{in assigning labels. In total, they discussed 7\% of the user stories, for which disagreement was observed. This reconciliation meeting lasted one hour and 20 minutes.} \rev{The names of the new categories introduced, which form the \textit{app-inspired categories}, were homogenised and consolidated in another meeting that lasted around 3 hours.} The consolidated set was finally assessed by the 1st author.

\smallskip
\noindent
\rev{To summarise, in both Phase I and II, the researchers categorised each individual user story. In Phase I, the categories could be original ones, and novel ones emerging from interview-based elicitation (\textit{interview categories}). In Phase II, the categories could be original categories, interview categories, and novel ones emerged during app store-inspired elicitation (\textit{app-inspired categories}). In Phase I, researchers were also asked to indicate whether a user story in an original category was something novel for that specific category (marked as R), or something already existing in that category (marked as E). This distinction between E and R was not adopted in Phase II, were only the category was specified. At this stage, it would not be practical to have such a fine-grained comparison, considering all the interview categories, plus the original categories---for a total of 25 categories.}




\subsection{Data Analysis}
\label{sec:dataanalysis}
Data analysis is carried out based on the results of the comparison activities, and on a thematic analysis of the selected apps and associated rationales provided by analysts.  

To answer \textbf{RQ1.1} and \textbf{RQ2.1}, we perform a quantitative statistical analysis, as these questions are specifically concerned with measures of differences. We thus consider the following study variables, oriented to give a quantitative representation of the concepts of initial customer ideas, documented requirements, and extended requirements, as well as their differences. 

The dependent variables of the first phase of the study are:

\begin{itemize}
 \item \textbf{Conservation rate:} rate of produced user stories that include content that can be traced directly to the original set of user stories. More formally, given a set of user stories produced by a certain analyst, let $e$ be the number of user stories that are marked as \textbf{E} by the researchers for some of the original categories, let $T$ be the total number of user stories for the analyst, the conservation rate $c$ is defined as $c = e / T$. 
\item \textbf{Refinement rate:} rate of produced user stories including content that can be mapped to the original categories, but that provide novel content within one or more of those categories. Formally, the refinement rate $r$ is $r = en / T$, where $en$ is the number of user stories marked as \textbf{R}.
\item \textbf{Novelty rate:} rate of produced user stories including content that is novel, and cannot be traced to existing categories.
Formally, the novelty rate is $v = n / T$, where $n$ is the number of user stories marked as \textbf{N}. \rev{These user stories belong to \textit{interview categories}.}

\end{itemize}

\rev{To understand the rates, consider the following case. An analyst wrote 50 user stories: 12 of them can be mapped to the original categories and do not add content with respect to the original user stories (marked as E); 30 can be mapped to the original categories, e.g., \textit{Communication}, but also add novel content, e.g., related to social media (marked as R); 8 introduce entirely new categories (marked as N), e.g., \textit{Company Information} or \textit{Usability}. In this case, the conservation rate is $24\%$, the refinement rate is $60\%$, and the novelty rate is $16\%$.}

For the roles, we have analogous variables,
Role Conservation, Refinement and Novelty Rate, defined respectively as: $c_\rho = e_\rho / T_\rho$, $r_\rho = en_\rho / T_\rho$, and $v_\rho = n_\rho / T_\rho$, where:
\begin{enumerate}
\item $T_\rho$ is the number of roles identified by the analyst;
\item $e_\rho, en_\rho, n_\rho$ are the roles that can be traced to the initial roles, the roles that express a refinement of the original ones, and the new roles, respectively. 
\end{enumerate} 

Other dependent variables are those related to the second phase of the study, in which the analysts took inspiration from the app stores to produce additional user stories:

\begin{itemize}

\item \textbf{App Store Novelty rate:} rate of user stories,  produced after the analysis of similar apps, which belong to novel \textit{categories} introduced in this phase by the analyst. Formally, the app store novelty rate is $v^{App} = a / S$, where $a$ is the number of user stories belonging to entirely novel categories for the specific analyst, and $S$ is the total number of user stories produced in this second phase by the analyst. This rate indicates to what extent the additional activity helped the analysts in identifying novel features not explored in the previous phases. \rev{Consider for example the following scenario. The user stories that the analyst $X$ defines after the interviews belong to 4 out of 5 original categories and to interview category ``Usability''. 
After the analysis of the apps, $X$ defines 20 user stories. 16 of them belong to the 4 original categories already used and to ``Usability''. The other user stories belong to the remaining original category, and to the category ``Availability'', which was introduced already by other analysts in the Phase I.  
In this case, $v^{App}$ is $20\%$ as there are 4 out of 20 user stories belonging to categories that $X$ did not explore in Phase I.}

\item \textbf{App Store Refinement rate:} this rate 
represents the rate of user stories that belong to the same categories already used by the analyst. Formally, this rate is $r^{App} = o / S$, where $o$ indicates the number of user stories belonging to existing categories used by the specific analyst\footnote{\rev{In some limited cases, user stories can belong to more than one category. So, while in principle $r^{App} = 1 - v^{App}$, in some cases $r^{App} + v^{App} > 1$.}}. This rate indicates to what extent the additional activity helped in refining the features identified in the previous phase.  
\rev{In the previous example, $r^{App}$ is $80\%$, as $16$ out of $20$ user stories belong to categories already considered by the analyst in Phase I.}
\end{itemize}

It should be noticed that in this phase we have no conservation rate, as we are not comparing with existing user stories. Concerning roles, we consider the app store role novelty rate as $v^{App}_\rho = a_\rho / S_\rho$, where $a_\rho$ is the number of new roles introduced in this phase for the specific analyst and $S_\rho$ is the total number of roles used by the analyst in the whole set of user stories. We do not consider other rates for roles, as we want to focus solely additional profiles introduced after app store-inspired elicitation.

Based on these rate variables, we want to see the interval, for which we can state that, for a confidence level of $0.95$ ($\alpha=0.05$), the rate variable $Y$ is comprised between a certain lower bound $L_Y$ and a certain upper bound $U_Y$. This can be achieved by identifying the \textit{confidence interval} of the mean of the sample of each rate variable $Y$---or the confidence interval of its median, when the data are not normally distributed. To test the normality assumption, we use the Shapiro-Wilk Test. When the assumption is met, we apply the one sample t-test. In the other cases, we apply the percentile method.
The labeling procedure and theme extraction performed by the 2nd and 3rd authors (Step 8 and 9 in Data Collection)  produced additional information that can help to answer \textbf{RQ1.2}, \textbf{RQ1.3}, as well as \textbf{RQ2.2} and \textbf{RQ2.3}. Indeed, these questions are concerned with the categories of user stories and roles, and their evolution in the different steps. 

\rev{
In particular, we are interested in analyzing the following indicators, with associated analysis. 
\begin{itemize}
\item \textbf{Categories and Roles Frequency}: frequency of user stories belonging to each category and role, considered as a proxy to evaluate their relevance. 
\begin{itemize}
    \item To answer \textbf{RQ1.2}, we compare the frequency in the original user stories, and the frequency in the user stories produced in Phase I. This comparison will be performed based on the original categories, and we will use one sample t-test, or Wilkoxon---if normality assumptions are not satisfied---with $\alpha=0.05$, to evaluate the differences. The goal will be to identify whether there is a difference between the original frequencies and the ones observed in the data produced after interviews. Additional statistics will also be provided to complement statistical tests.
    \item To answer \textbf{RQ2.2}, we compare the frequencies in the user stories produced in Phase I with the frequencies in the user stories produced in Phase II by the same analysts. The comparison will be performed  considering \textit{original}, \textit{interview}, and \textit{app-inspired} categories and roles. We will use paired t-test, or Wilkoxon---if normality assumptions are not satisfied---with $\alpha=0.05$, to evaluate the differences. The goal is to understand whether there is a difference between the distributions of the different categories in Phase I and in Phase II. This helps to  understand if the focus of the analysts changed in the two phases. Additional statistics will also be provided to complement statistical tests.  
\end{itemize}
\item \textbf{Emerging Categories and Roles}: specific names and associated frequencies of interview categories and roles (to answer \textbf{RQ1.3}) and app-inspired categories and roles (to answer \textbf{RQ2.3}). 
\end{itemize}

Given the \textit{exploratory} nature of the study, we do not make an explicit definition of all the variables involved, and all the hypotheses tested throughout the evaluation, which would suggest a confirmatory study design. Instead, in the following results sections, we will focus on presenting multiple statistics to provide evidence of the evolution of requirements throughout the different elicitation phases. 
}




%% file: sections/results.tex
\section{RQ1: Interview-based Elicitation---Execution and Results}
\label{sec:rq1results}

In this section, we report the results of the activities related to interview-based elicitation, evaluating the confidence intervals for the different rates (\textbf{RQ1.1}), the variations in terms of distributions of categories and roles with respect to initial ideas (\textbf{RQ1.2}), and the novel categories and roles introduced by the analysts (\textbf{RQ1.3}).

\subsection{RQ1.1: How \rev{large} is the difference in terms of documented requirements and roles with respect to initial ideas?}



\begin{figure}[t]
\centering
\includegraphics[width=\columnwidth]{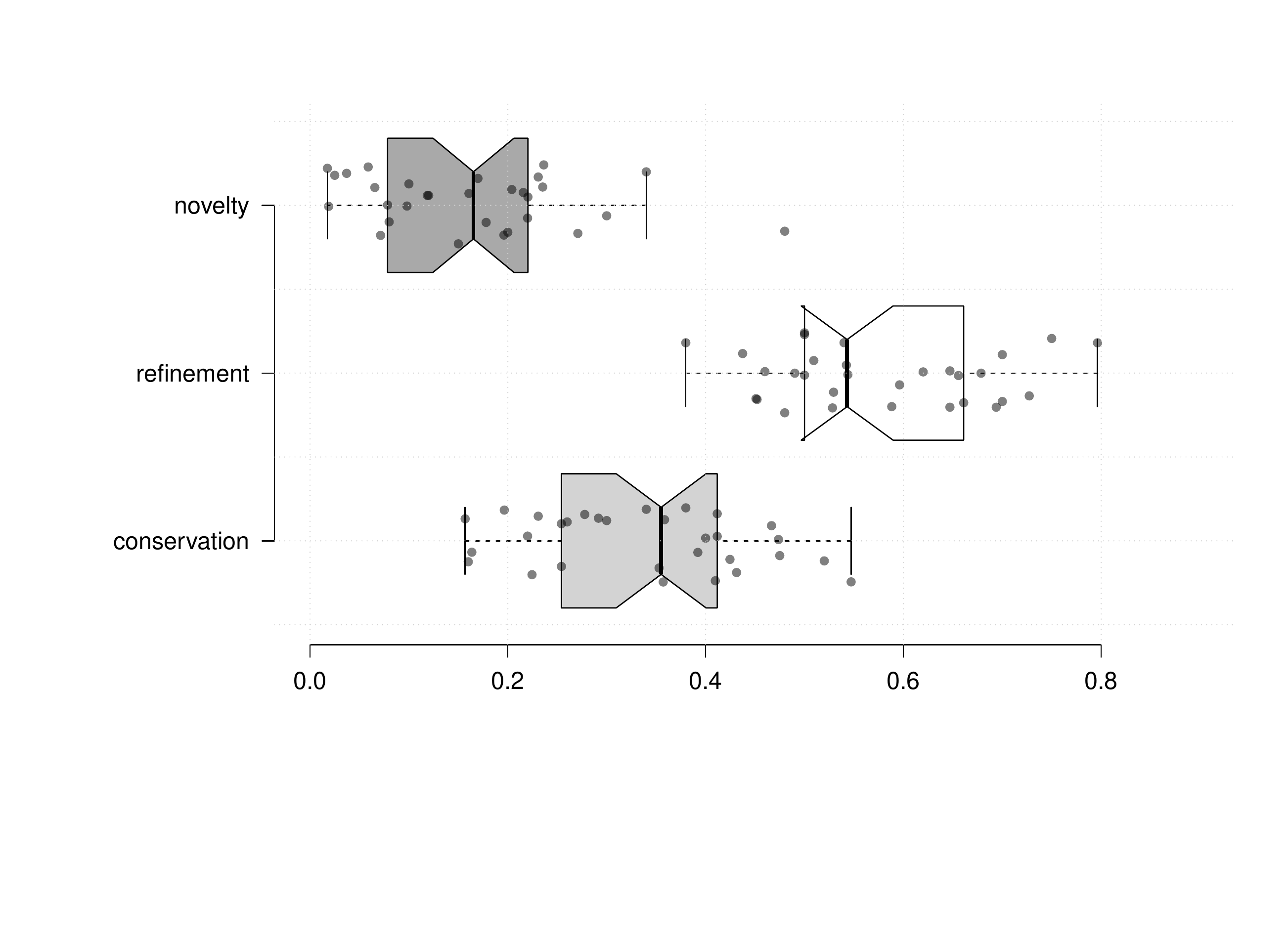}
\caption{Boxplots of the different rates \rev{(cf. Sect.~\ref{sec:dataanalysis} for definitions)}: \rev{\textbf{conservation} = conservation rate; \textbf{refinement} = refinement rate; \textbf{novelty} = novelty rate.} The non-overlapping notches indicate that the difference between the median of the rates is significant for a confidence level of 95\%. \rev{From the figure, we  observe that most of the user stories produced are refinements, followed by existing user stories and by novel ones.}}
\label{fig:ratesboxplot}
\end{figure}

\begin{figure}[t]
\centering
\includegraphics[width=\columnwidth]{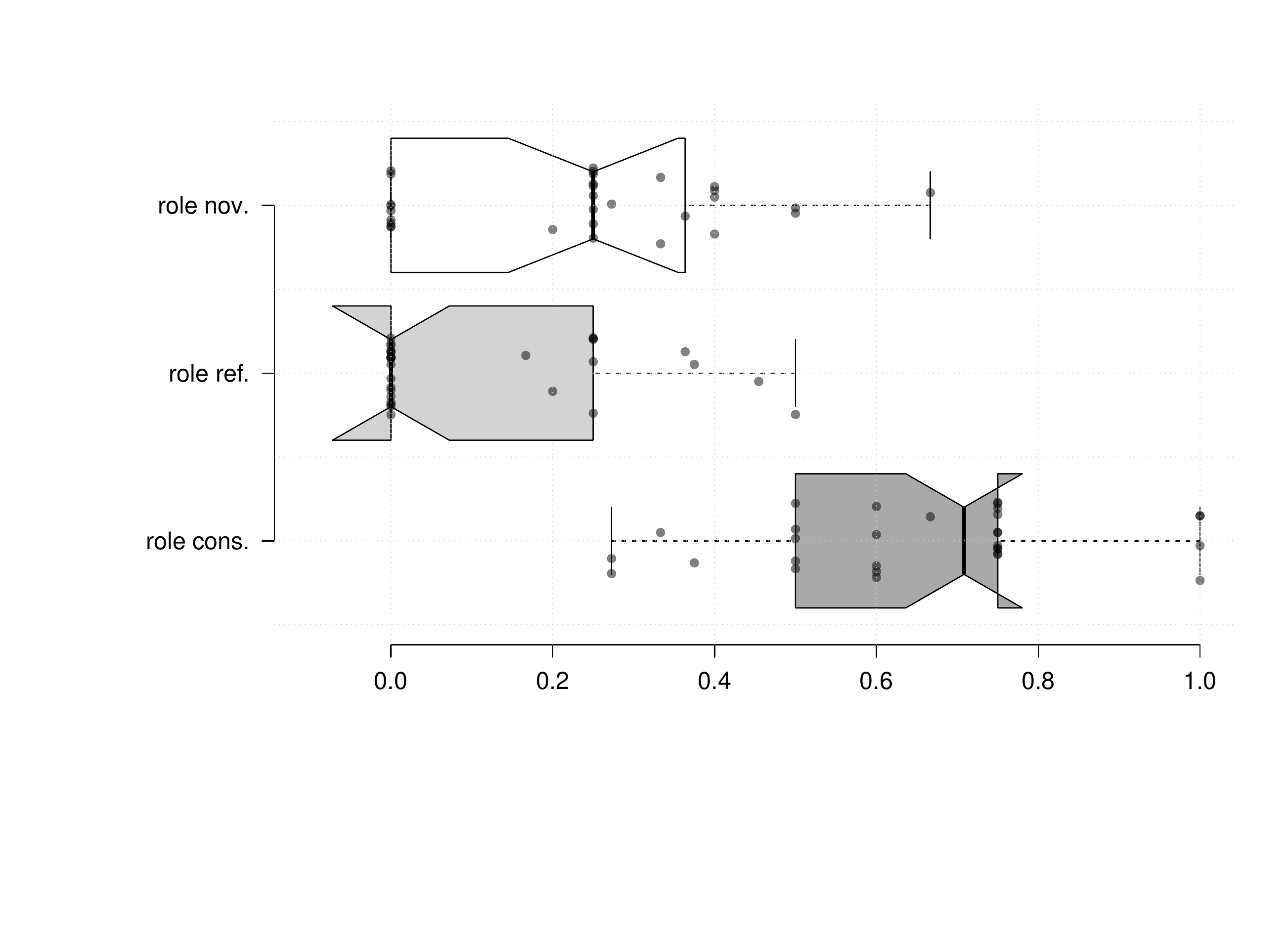}
\caption{Boxplots of the different rates. \rev{(cf. Sect.~\ref{sec:dataanalysis} for definitions)}: \rev{\textbf{role cons.} = role conservation rate; \textbf{role ref.} = role refinement rate; \textbf{role nov.} = role novelty rate.} The non-overlapping notches between role conservation and the other two rates indicate that the difference between the median is significant, with a confidence level of 95\%. Instead, the difference between role refinement and novelty is not significant. \rev{From the figure, we also observe that most of the roles belong to the initial set (role cons. is the highest), while novel ones or refinements are less frequent.}}
\label{fig:roleratesboxplot}
\end{figure}

\begin{figure}[t]
\centering
\includegraphics[width=\columnwidth]{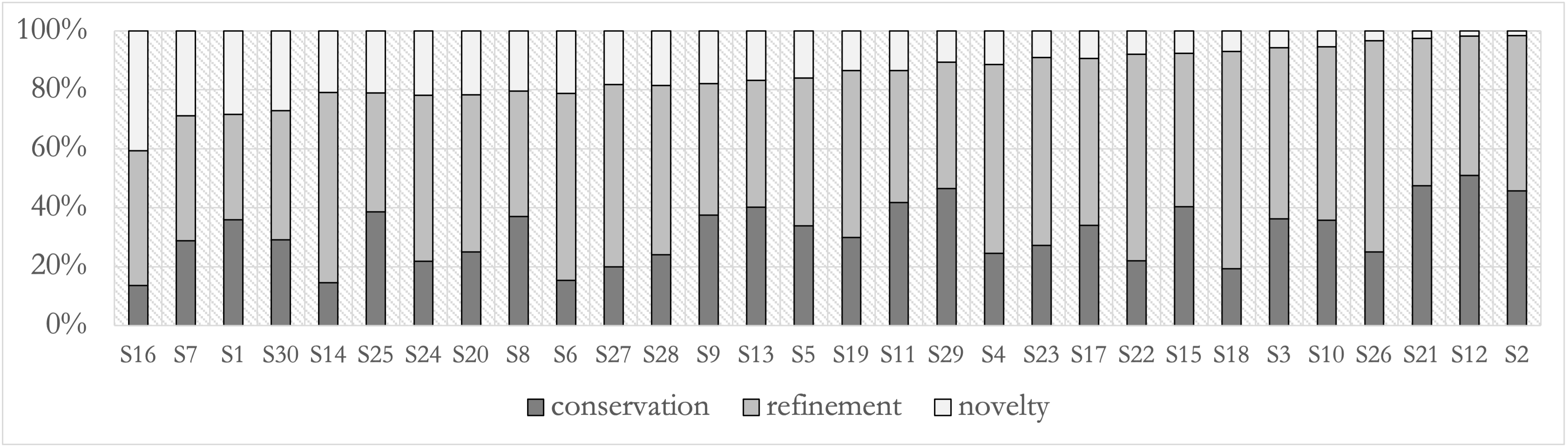}
\caption{Relative percentage of the different rates for each analyst.}
\label{fig:rates}
\end{figure}

\begin{figure}[t]
\centering
\includegraphics[width=\columnwidth]{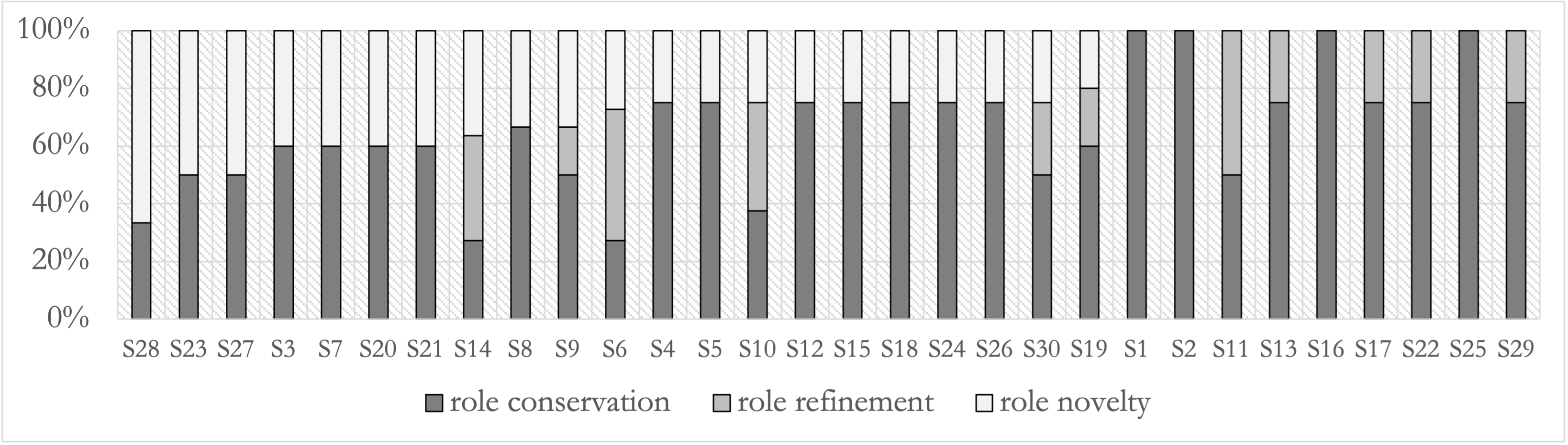}
\caption{Relative percentage the different role rates for each analyst.}
\label{fig:rolerates}
\end{figure}

In Fig.~\ref{fig:rates} and Fig.~\ref{fig:rolerates}, we report the plots of the values of the different rates for each analyst. 
Instead, the boxplots in Fig.~\ref{fig:ratesboxplot} and~\ref{fig:roleratesboxplot} give an overview of the distribution of the rates. We see that in general the highest values are observed for the refinement rate, followed by the conservation rate and by the novelty rate. Conversely, for roles, conservation rate dominates over novelty and refinement ones. 
Looking at Fig.~\ref{fig:rates} and Fig.~\ref{fig:rolerates}, we intuitively see that variations for each rate are quite high from an analyst to the other. This suggests that each individual analyst produces different user stories in terms of content, and thus, depending on the analyst, different systems may be developed. In some cases, analysts lean more towards the refinement of user stories in the existing categories, while in others focus on completely novel features, as one can observe, e.g., for analysts 7 and 16. In other cases, e.g., analysts 2 or 12, the elicitation process tends to be more conservative, with less space for creativity.

In Table~\ref{tab:ratetest}, we answer RQ1 by identifying the confidence intervals for each rate variable. 

\input{tables/tab-test-rates}

Based on the tests results, the following statements can be given, for a confidence level of 95\%:

\begin{itemize}
  \item The conservation rate $c$ is between $30$\% and $38$\%, meaning that between $30$\% and $38$\%  of the produced user stories \rev{(i.e, roughly up to 19 out of 50)} can be fully traced to user stories belonging to the initial ideas.
  \item The refinement rate $r$ is between $54$\% and $62$\%, meaning that between $54$\% and $62$\%  of the produced user stories \rev{(i.e, roughly up to 31 out of 50)} are refinements of initial ideas. Specifically, they belong to categories of features already conceived by the customer, but provide novel content. 
  \item The novelty rate $n$ is between $12$\% and $20$\%, meaning that between $12$\% and $20$\%  of the produced user stories \rev{(i.e, roughly up to 10 out of 50)} identify entirely novel categories of features, which did not belong to the initial ideas.  
  \item The role conservation rate $c_\rho$ is between $58\%$ and $72$\%, meaning that between  $58\%$ and $72$\% of the roles used in the produced user stories match with the initial roles identified by the customer. 
  \item The role refinement rate $r_\rho$ is between $6\%$ and $17\%$, meaning that between $6\%$ and $17\%$ of the roles used in the produced user stories are refinements of the initial roles. 
  \item The role novelty rate $n_\rho$ is between $18\%$ and $29\%$, meaning that between $18\%$ and $29\%$ of the roles are entirely novel with respect to the initial ones. 
\end{itemize}

\subsection{RQ1.2: What is the relevance given to the categories of requirements and roles with respect to initial ideas?}

To answer RQ1.2, we analyze how the relevance given to each category and each role change in the analysts' stories with respect to the original ones. \rev{We remark that, with the term ``relevance'' we arguably intend the fraction of the stories dedicated to a certain category or role.}
This analysis will allow us to identify what is important for the analysts and to reflect on the meaning of these preferences.

\subsubsection{\textbf{Analysis of Categories}}
\label{sec:originalcategories}
Through the process described in Section~\ref{sec:datacollection}, the researchers identified 5 different categories:
\begin{itemize}
\item \textbf{Administrative procedure related to customers} (labeled as \textit{customers}): this category includes features such as registration to a camp, creation of new campers and parents profiles, modification, and elimination of profiles.
\item \textbf{Management of facilities} (\textit{facilities}): this category includes the tracking of the facilities' status both in terms of usage and maintenance needs and the management of the inventory.
\item \textbf{Administrative procedure related to personnel} (\textit{personnel}): This category includes features related to assigning tasks to workers and evaluating them. In its more general interpretation, it can also include the ability to create and modify personnel profiles and other administrative needs.
\item \textbf{Individual camp management} (\textit{camp}): This category includes features such as scheduling activities within a specific camp, managing its participants, and dealing with additional planning details.
\item \textbf{Communication} (\textit{communication}): This category includes everything related to communication from one-to-one messaging and broadcasting to a specific category of users to posting information online and in social media.
\end{itemize}

\paragraph{Initial Categories Distribution}
In the original stories, and, thus, in the mind of the fictional customer, great relevance is given to the administrative activities on the customer side (i.e., stories belonging to \textit{customers}), such as ``As a camp administrator, I want to be able to add campers so that I can keep track of each individual camper''. 
In particular, $64.2\%$ of the total number of stories belong to this category, and the remaining is distributed among the other categories as follows: 5.7\% belong to \textit{facilities}, 9.4\% to \textit{personnel}, 7.6\% to \emph{camp}, and 15.1\% to \textit{communication}. 

\paragraph{Categories Distribution in the Analysts' Stories}
\input{tables/descriptive-categories}

The category distribution of the analysts' stories is different with respect to the original stories. Indeed, looking at Table~\ref{tab:categoryDistribution}, we observe that the mean of the percentage of stories belonging to \emph{customers} is lower than half of the one in the original stories, while the mean of \emph{communication} is double of the value for the original stories. 

\begin{figure}[t]
\centering
\includegraphics[width=\columnwidth]{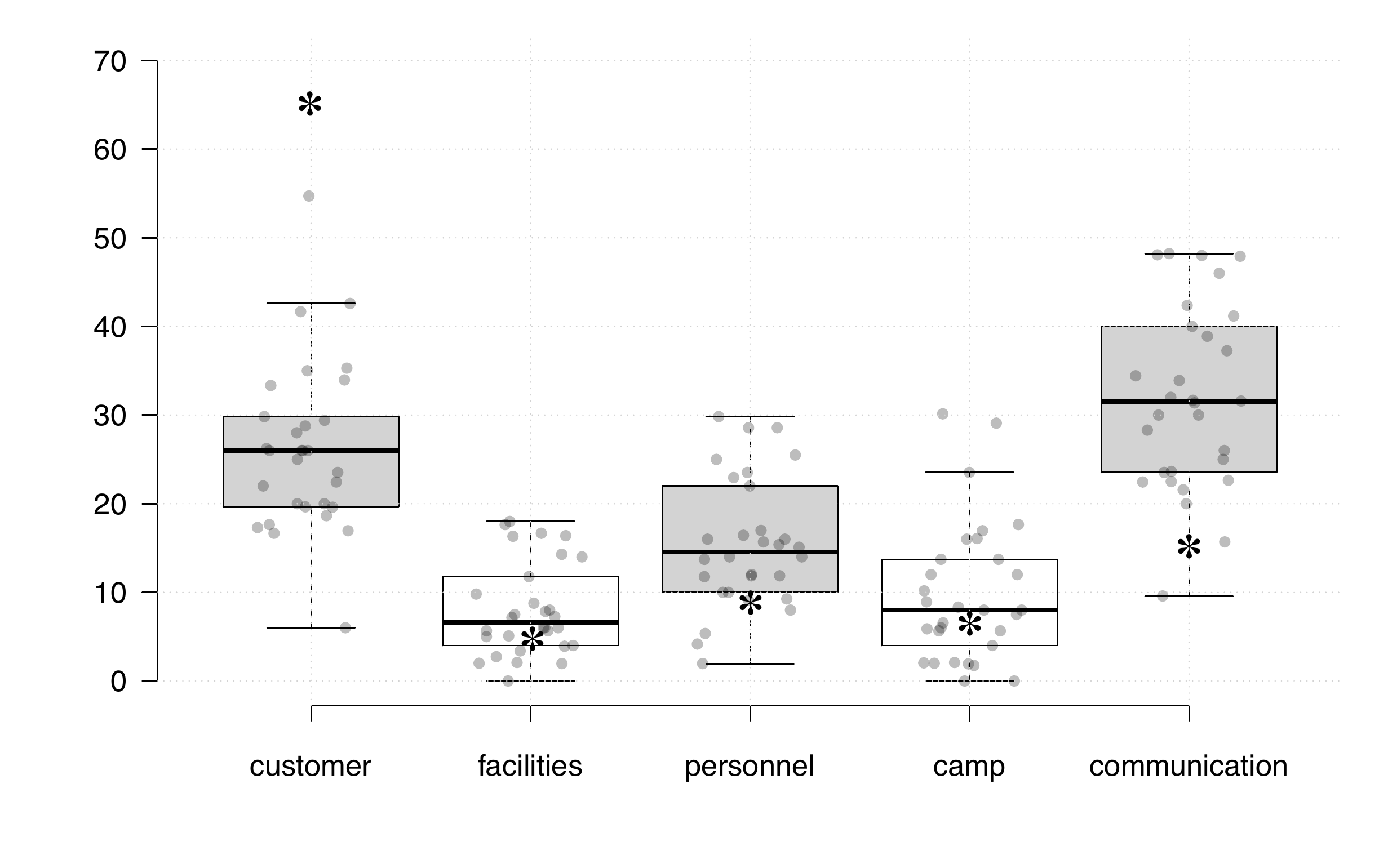}
\caption{Box plots of the distributions for each category, measured as the percentage of user stories in the category. The black asterisks * represent the percentage in the original set. \rev{The figure shows higher interest in communication-related user stories with respect to the original set, and much less interest in customer-related user stories.}}
\label{fig:customerFreq}
\end{figure}

\input{tables/test-initial-cats}

The box plots in Fig.~\ref{fig:customerFreq}, in which the black asterisks represent the percentage for each category in the original stories, show that the relevance given to \textit{facilities}, \textit{personnel}, and \textit{camp} of the original stories is comparable with the one given by the analysts. Instead, for \textit{customers} and \textit{communication} there are strong differences. 
This suggests that the analysts focused their attention on aspects that were not the original focus of their customer. Table~\ref{tab:testoriginalcats} reports the statistical tests to assess whether there is a difference between the means of the distribution and the relevance in the original user stories. The tests show that the differences between observed mean (Obs $\mu$) and expected one (Exp. $\mu$) are significant in all the cases for $\alpha = 0.05$. The tests are all one-sample tests. We use the t-test when the normality assumption is fulfilled, and Wilcoxon Signed Rank in the other cases. It is rather striking to observe that the decrease of relevance for the customer category is about 59\% (i.e., 37.81).

\subsubsection{\textbf{Analysis of Roles}}
Similar considerations can be done about the perspective considered in imagining the system, and, thus, the roles used in the user stories. 
In the original set of user stories, there are three roles: \textit{camp administrator}, \textit{parent}, and \textit{camp worker}. 

\paragraph{Initial Roles Distribution}
The majority of stories is dedicated to the camp administrator (66.0\% of the user stories), followed by the parents of the participants (24.5\%) and the camp worker (9.4\%).
This also helped the preparation of the fictional customer who was acting as the camp administrator as he had most of the available stories looking at the system to be developed from his perspective. 

\paragraph{Roles Distribution in the Analysts' Stories}
The analysts had only the chance to talk with the administrator and, nevertheless, 30\% of them in their stories were dedicated to other roles, and only 16.67\% of them had more than half of the stories focused on the administrator.

\input{tables/descriptive-roles}

Table~\ref{tab:roleDistribution} shows the descriptive statistics for the distribution of the analysts' user stories among the different roles. 
It is interesting to observe that the mean for camp workers is much higher than the one on parents. This could be connected to the decreasing attention to the category \textit{customers} in the analysts' requirements. 
Notice that, while all the analysts considered the role of camp administrator, two of them did not consider the role of camp workers, and four did not consider the role of parents. 

\begin{figure}[t]
\centering
\includegraphics[width=\columnwidth]{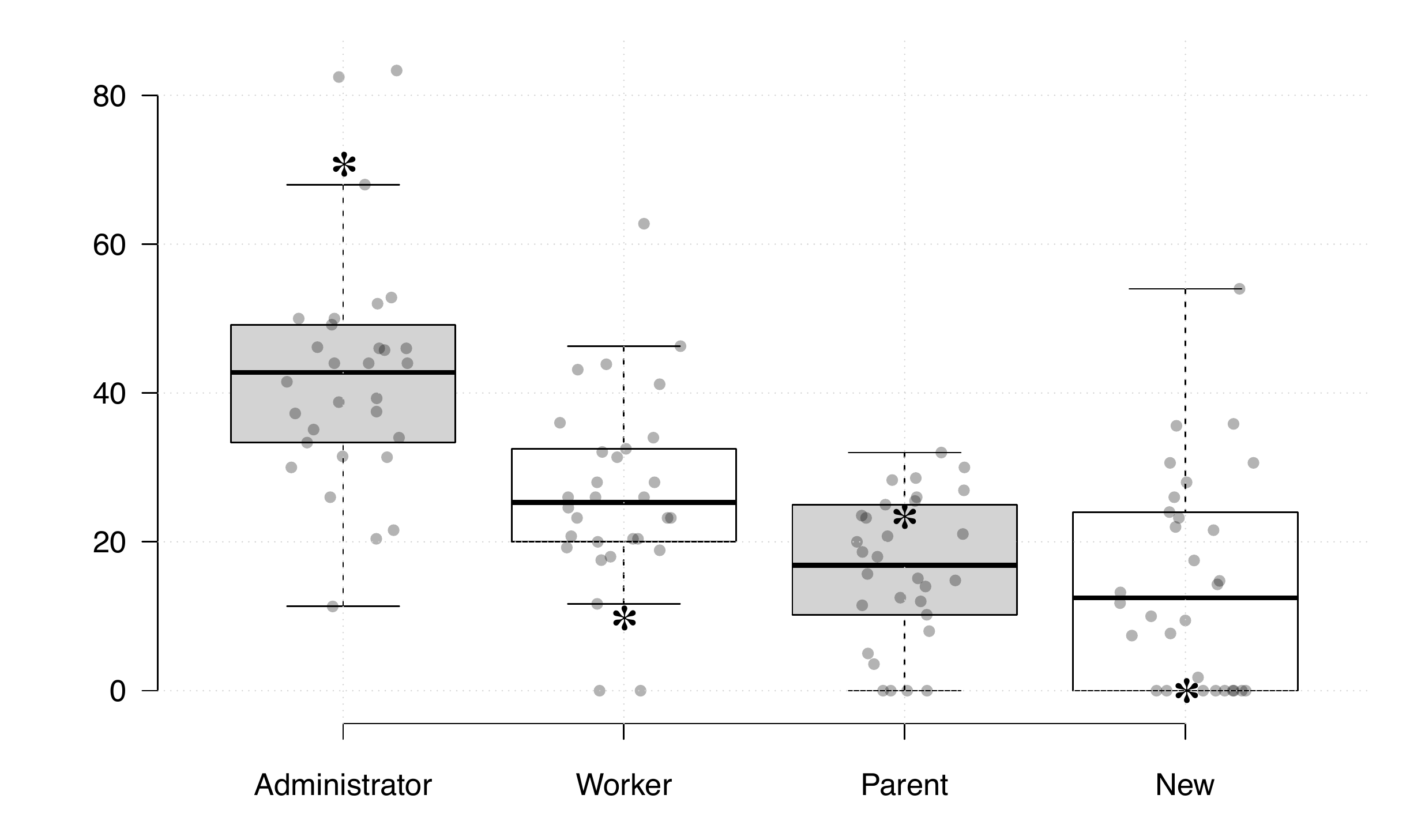}
\caption{Box plots of user story distribution for each role, measured as the percentage of user stories that used that role. The asterisks * are the number of user stories for the specific role in the original set. \rev{The figure shows higher interest in the role of \textit{Workers} with respect to the original set, and much less interest in the \textit{Administrator} role, which is also the customer that has been interviewed.}}
\label{fig:roleFreq}
\end{figure}

As shown in the box plots in Fig.~\ref{fig:roleFreq}, in  which  the  black  asterisks represent  the  percentage  for  each  role  in  the  original stories, only the outlier focused more on the camp administrator role than the original stories. Similarly, the relevance given to the parents' perspective is in general much smaller than in the original stories. 

\input{tables/test-roles-phase-1}

The impact of new roles is limited (mean of 13.8\%).  All the analysts considered at least 2 of the original roles, and 80\% of them considered all the 3 original roles. However, the relevance given to the perspectives and roles in the analysts' user stories is considerably different than in the original ones. This is confirmed by the statistical tests in Table~\ref{tab:testrolesphase1}, which show that the the percentage of the original stories for each role is significantly different from the one observed in our data. 

\subsection{RQ1.3: What are the emerging categories and roles?}

To answer RQ1.3, we analyze the new categories and new roles, their recurrence among analysts, and their weight within the set of stories of the analysts who included them. 
This analysis provides insight on what is generated by the analysts' expertise, background, preparation, and analysis. 

\subsubsection{\textbf{Analysis of New Categories}}

\begin{figure}[h]
\centering
\includegraphics[width=\columnwidth]{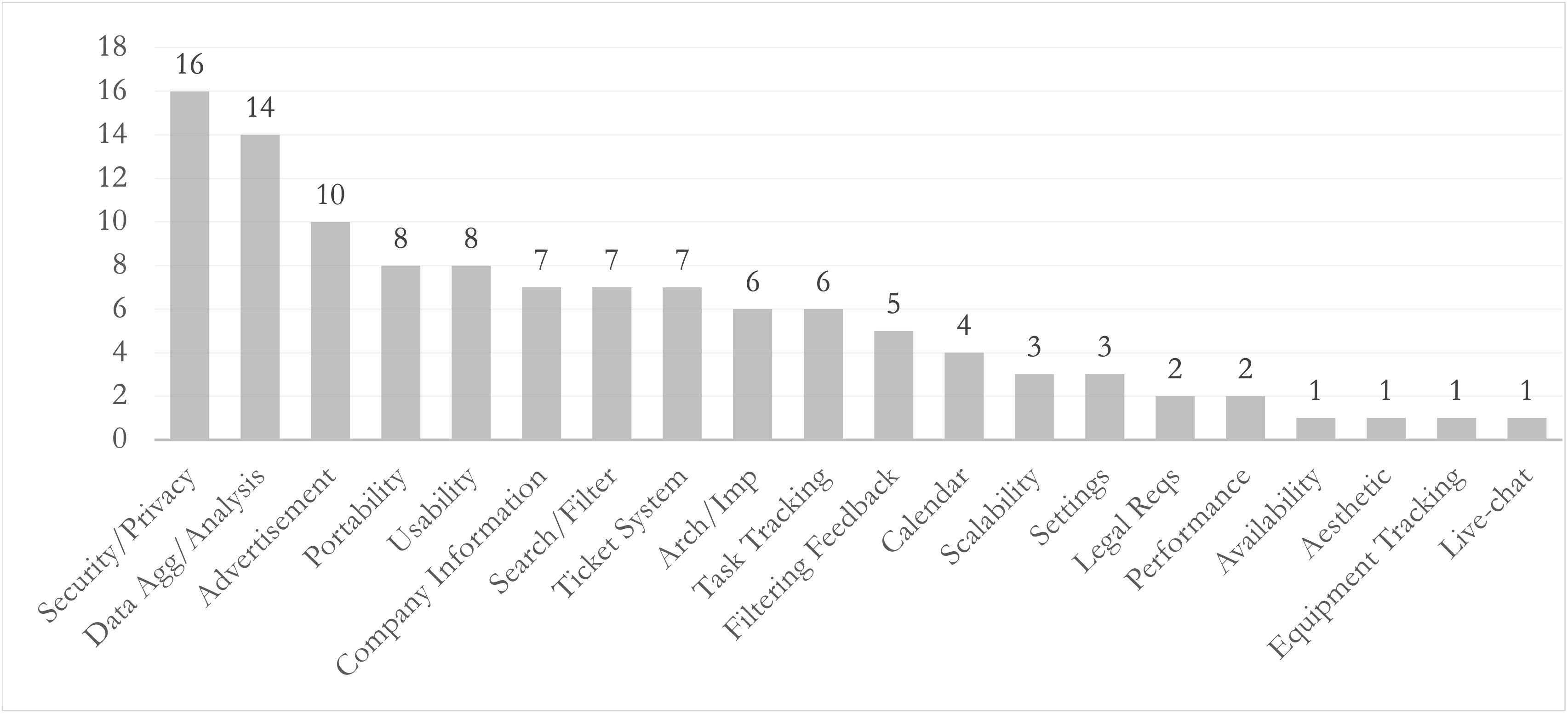}
\caption{Distribution of the novel categories that emerged after interview-based elicitation. The histogram reports the number of analysts who considered the specific category.}
\label{fig:newCategories}
\end{figure}

In their stories, every analyst included new categories up to a maximum of 8 with a mean of 3.73. In total, they introduced 20 new categories, reported in Fig.~\ref{fig:newCategories}. Among them, 4 have been used by a single analyst but were very specific and could not map over any other existing or newly created categories.

Table~\ref{tab:NewCat} reports descriptive statistics on new categories. The more recurrent new category, used by 53\% of the analysts, is \textit{Security/Privacy}. The impact on the total of user stories in this category varies from 1.79\%, one story, to 10\%, 5 stories. An example of such a story is ``As an employee, I want this app to not have access to my personal phone data so that what I have on there stays private''.

\input{tables/descriptive-newCat}

The second more recurrent new category is \textit{Data Aggregation/Analysis}. This category collect all the features that aim at aggregating and analyzing information at company-level. An example is ``As a business owner, I want to be able to analyze the data that is entered into the system so that I can see trends in the information that I receive.''.
The mean usage is 3.9\% (2 stories) with std. dev. 2.1\% as the relevance given varies from 1.9\% to 7.8\%. 

Notice that there is a fundamental difference between this category and the previous one. ``Security/Privacy'' includes nonfunctional requirements that many of the analysts decided to investigate during their conversation with the customer, while ``Data Aggregation/Analysis'' includes functional requirements that describe global operations at a company level.

The third more recurrent new category is \textit{Advertisement}, used once or twice by almost a third of the analysts. This category includes the functionalities \rev{that} the analysts identified to advertise the camps to potential guests and their parents. An example is ``As a camp administrator, I want to have the ability to advertise scheduled activities so that potential camp attendees and their guardians will know what activities are upcoming''.

Among the other nonfunctional requirements categories that emerged in the analysis are \textit{Usability} and \textit{Portability}, which have both been considered by 8 participants (26.7\% of the total). 
An example that belongs to the \textit{Usability} category is ``As an employee, I wish for this app to be simple to use so that our older staffers can still use it''. Notice that, when present, usability is highly considered with an average of 6.9\% of the stories. \textit{Portability} has a lower impact on the stories of the analysts who use it (4.2\%). An example in this category is ``As a camp administrator, I want users to be able to use the system on all platforms so that no users are excluded from seeing what the campsites have to offer.''.

The functionalities related to publishing and accessing to general information about the companies (e.g., its contact information) were not present in the original user stories and have been collected in the category \textit{Company Information}. This category has also been considered by 7 participants.


In synthesis, we observe that new categories emerged for every analyst and some of them are highly predominant both in terms of number of analysts who considered them and in impact on the stories of the analysts who considered them. The new categories are almost equally divided into new functionalities and nonfunctional requirements.

\subsubsection{\textbf{Analysis of New Roles}}
Differently from the case of categories, not all the analysts added new roles to their stories, but still a \rev{large number} did it (\rev{22 analysts over 30}).
6 new roles emerged from the analysis, namely, in order of frequency, attendee (16), visitor (8), user (7)\footnote{User can be regarded as a collective, abstract role.}, consultant (3), system administrator (2), and investor (1). 
Table~\ref{tab:NewRole} reports descriptive statistics on new roles (excluding system administrators and investors that appear just once in the stories of 2 and 1 analysts, respectively). The most frequent among these roles, attendee, refers to the children participating in the camps and thus assumes that they all will have access to the system---while this was not possible in the original system's idea. When used, this role has a high impact with a mean of 15.8\% (standard deviation of 8.4\%), \rev{meaning that analysts who introduced this role, used it in 15.8\% of their stories}.  Notice that the analyst who used it more dedicated 38\% of the stories to this role, which represented the most frequent role in the analyst's set. Summarizing, we observe that many analysts consider additional roles with respect to the ones in the original set.

\input{tables/descriptive-NewRoles}


\section{RQ2: App-Store Inspired Elicitation--- Execution and Results}
\label{sec:rq2results}

In this section, we report the results of the activities related to app store-inspired elicitation, evaluating the confidence intervals for the different rates (\textbf{RQ2.1}), the variations in terms of distributions of categories and roles with respect to the previous phase (\textbf{RQ2.2}), and the novel categories and roles introduced (\textbf{RQ2.3}).

\subsection{RQ2.1: How \rev{large} is the difference in terms of covered requirements categories  and  roles  with  respect  to  the  requirements  documented  after  the interview  sessions?}

\begin{figure}[t]
\centering
\includegraphics[width=\columnwidth]{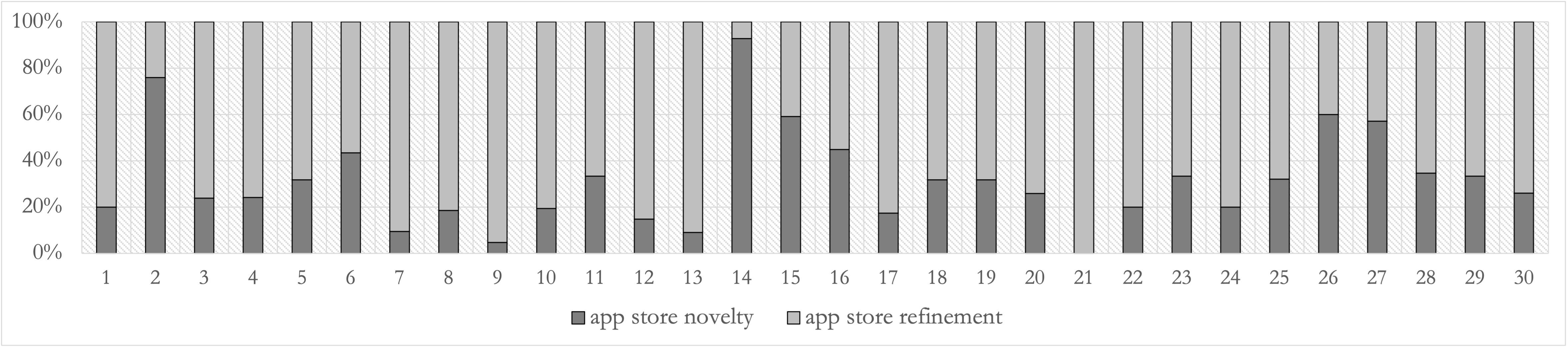}
\caption{Relative percentage of app store novelty and refinement rates.}
\label{fig:appstorechart}
\end{figure}

\begin{figure}[t]
\centering
\includegraphics[width=\columnwidth]{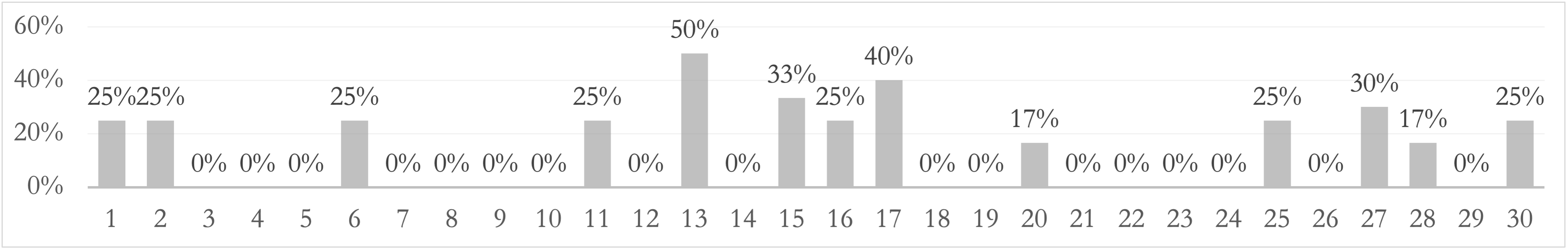}
\caption{App store role novelty rate for each analyst, expressed as percentage.}
\label{fig:appstorerolechart}
\end{figure}

\begin{figure}[t]
\centering
\includegraphics[width=\columnwidth]{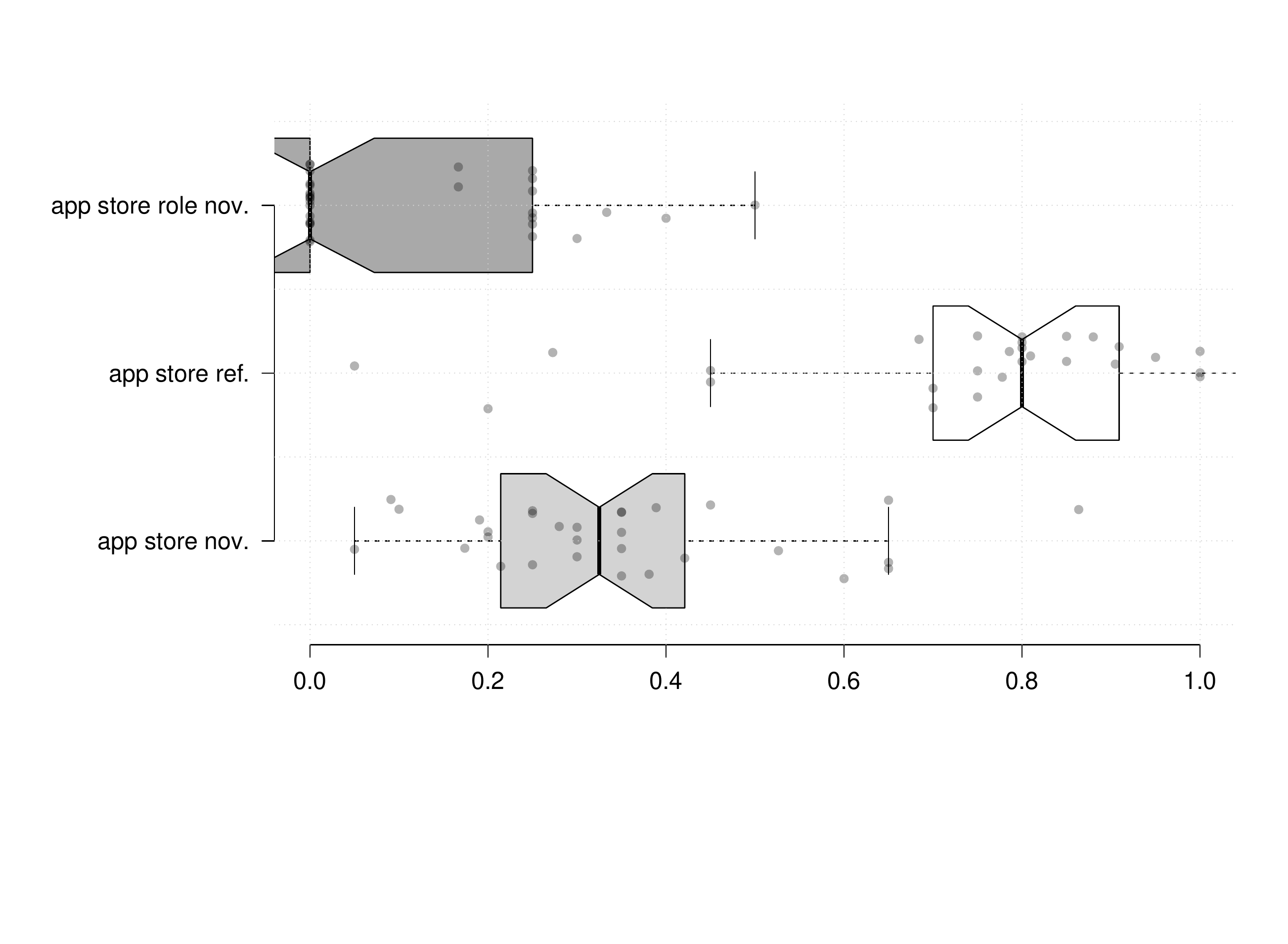}
\caption{Boxplots of the different app store-related rates. \rev{(cf. Sect.~\ref{sec:dataanalysis} for definitions)}: \rev{\textbf{app store nov.} = app store novelty rate; \textbf{app store ref.} = app store refinement rate; \textbf{app store role nov.} = app store role novelty rate.} The figure shows that most of the user stories belong to categories already considered (app store ref. is higher than app store nov.). Furthermore, the novel roles are generally more limited with respect to those already discovered---app store role nov. is below 0.3.}
\label{fig:appstoreboxplot}
\end{figure}

Fig.~\ref{fig:appstorechart} reports the relative percentages of app store novelty and app store refinement rates for each analyst. We see that again there are several differences between analysts, with some of them (e.g., 2, 15) including a large percentage of novel categories, and some of them (e.g., 7, 9), sticking to the original ones. However, we notice all analysts included at least some novel categories in their user stories. For the role novelty rate, shown in Fig.~\ref{fig:appstorerolechart}, we see an even higher variability in the results, with some analysts (13, 15, 17) introducing different roles, and the majority of them (17 subjects out of 30) keeping the originally identified ones. Fig.~\ref{fig:appstoreboxplot} reports the boxplots of the rates, further highlighting the high variance of the app store role novelty rate. The non-overlapping notches indicate that the difference between the medians of the rates is significant for a confidence level of 95\%. 

Table~\ref{tab:testratesapp} reports the results of the tests performed to identify upper and lower bounds of the different rates. From the tests, we conclude the following, with a confidence level of 95\%:

\begin{itemize}
    \item The app store novelty rate $v^{App}$ is between 28\% and 42\%. This means that up to 42\% of the user stories produced after the analysis of similar apps \rev{(i.e., roughly up to 8 out of 20)} belong to categories that were not identified by the analyst after the first phases of elicitation.
    \item The app store refinement rate $r^{App}$ is between 68\% and 87\%, meaning that the majority of the user stories produced after the analysis of similar apps \rev{(i.e., roughly up to 17 out of 20)} are refinements of ideas already identified during the previous elicitation activities.
    \item The app store role novelty rate $v^{App}_\rho$ is between 8\% and 17\%, meaning that few additional roles are identified after app store-inspired elicitation by the individual analysts, but still a non-negligible number. 
\end{itemize}

\input{tables/tab-test-rates-apps}

\subsection{RQ2.2: What is the relevance given to the different categories of requirements  and  roles  with  respect  to  the  requirements documented after  the interview  sessions?}

To answer this RQ, we first look at the relevance of the categories by group, and then we look at the variation of single specific categories. The same will be done for roles.

\subsection{Analysis of Categories}

\begin{figure}[t]
\centering
\includegraphics[width=\columnwidth]{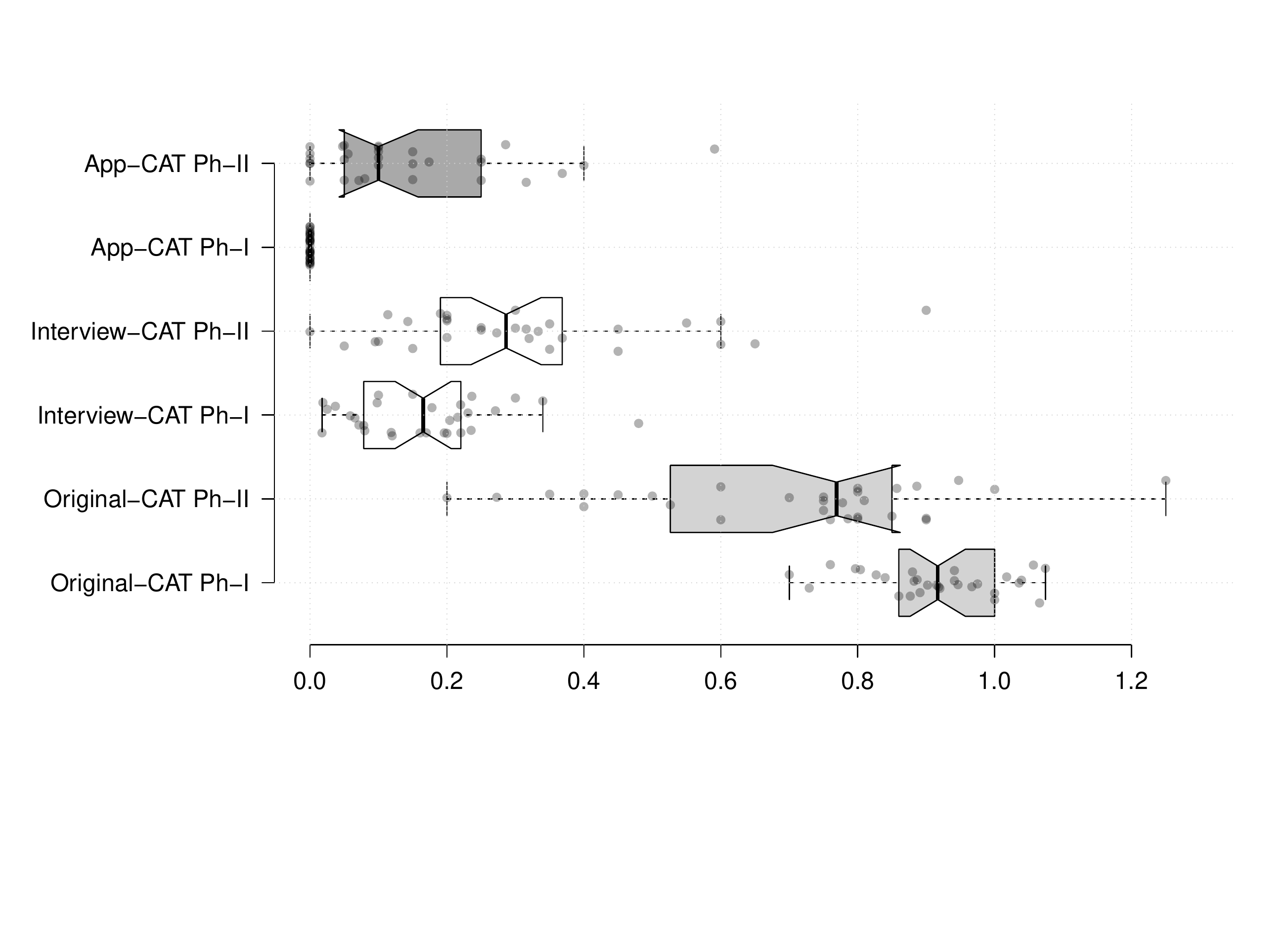}
\caption{Boxplots that represent the \rev{fraction} of user stories in each category group for the different phases, identified as Ph-I and P-II. \rev{The category groups are: \textbf{Original-CAT =} initial set of categories; \textbf{Interview-CAT =} new categories introduced in the interview phase; \textbf{App-CAT =} new categories introduced after the app store-inspired phase. The boxplots show that the original categories still dominate in Phase II. However there is less attention to them with respect to Phase I (Original-CAT Ph-I is lower than Original-CAT Ph-II), and increased attention to categories introduced in the interview phase  (Interview-CAT Ph-II is higher than Interview CAT Ph-I). Categories introduced in Phase II (i.e., App-CAT Ph. II) are the less frequent in Phase II, and do not obviously exist in Phase I.}}
\label{fig:boxplotscategoryshift}
\end{figure}

To facilitate the analysis, we group the categories into three groups, namely: Original, i.e., the set of categories that were part of the initial ideas; Interview-based, i.e., the categories that emerged after interview-based elicitation; App store-inspired, i.e., the categories that emerged after app store-inspired elicitation. Fig.~\ref{fig:boxplotscategoryshift} shows the boxplots of the distribution of the three category groups (abbreviated as \rev{Original-CAT, Interview-CAT and App-CAT}) in the two phases, identified as Ph-I and Ph-II. From the plot, we clearly see that the original categories received substantially less attention in Ph-II. Conversely, more relevance was given to categories in the Interview-based group. 
Overall, the novel categories belonging to the App group received less attention than the others, but still a non-negligible rate of user stories (mean 14\%) was dedicated to them.

The statistical tests reported in Table~\ref{tab:threegroupstest} show that the difference between the means of \rev{Original-CAT Ph-I and Original-CAT Ph-II, and Interview-CAT Ph-I and Interview-CAT Ph-II}, are statistically significant for $\alpha = 0.05$. Similarly, the variation from zero within the App category is also statistically significant. Therefore, we can conclude that in the different elicitation phases the analysts give relevance to different category groups. More specifically, we see: a mean decrease of 0.21 for the group Original, which is 23\% of the mean in Phase I (0.92); a mean increase of 0.14 for the group Interview, which is 45\% of the mean in Phase II (0.31).  Furthermore 14\% of the user stories in Phase II consider novel categories (group App, mean = 0.14).

\input{tables/tests-three-groups}

\begin{figure}[t]
\centering
\includegraphics[width=\columnwidth]{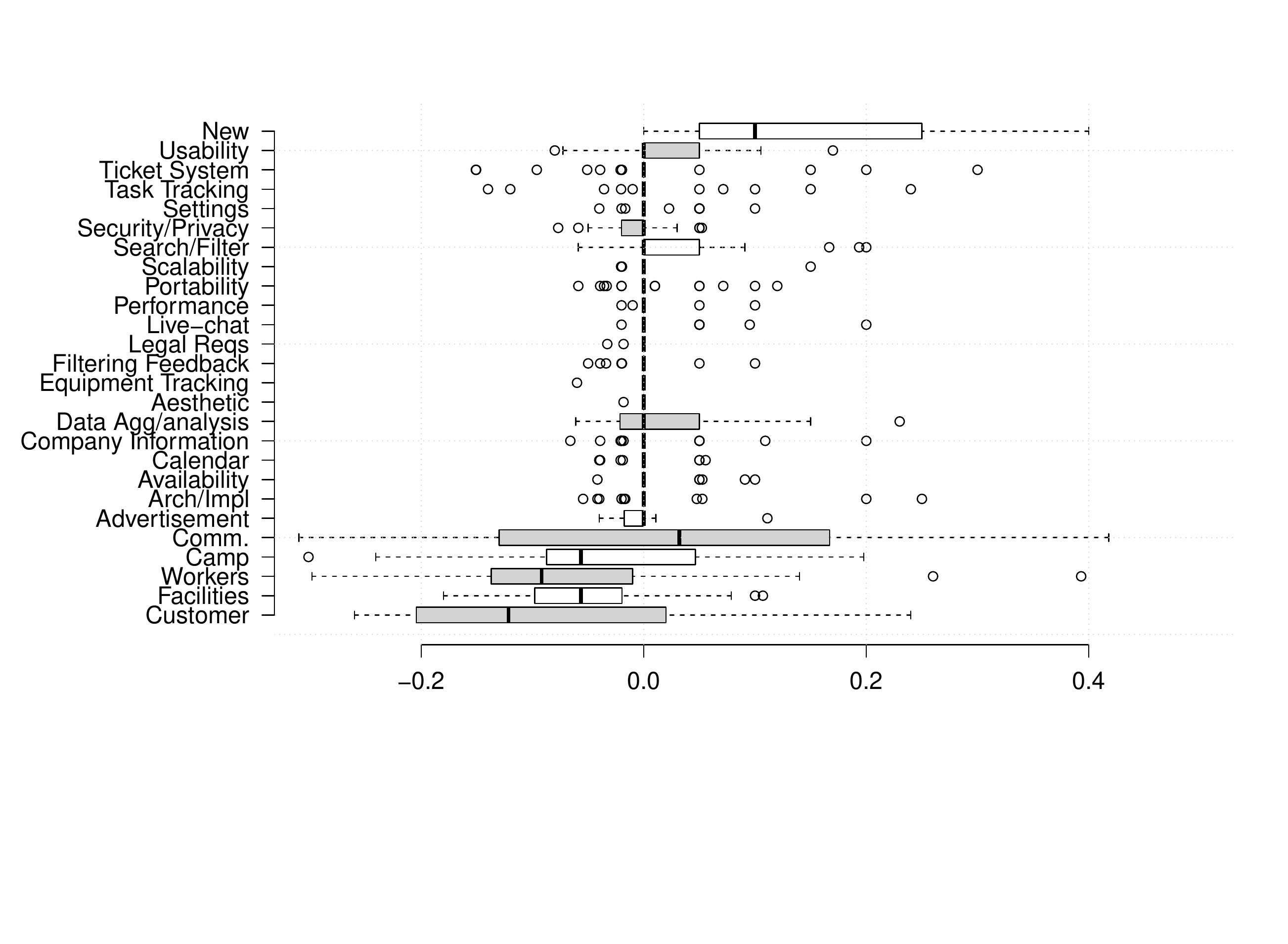}
\caption{Boxplots of the difference between the rate of user stories after interviews, and after app store-inspired elicitation for each category. \rev{The boxplot shows the  decrease of the original categories, i.e., the five ones reported at the bottom of the plot, with the exception of \textit{Communication} (Comm.)}.}
\label{fig:increaseboxplot}
\end{figure}

To look into the single categories, we consider, for each analyst, their individual rate of user stories in a certain category during Phase I and during Phase II.  Then we compute the absolute difference, to check whether some variation occurred in that category. The boxplots of the differences are reported in Fig.~\ref{fig:increaseboxplot}. We see that there are several categories with a limited number of data points, indicating that only a part of the analysts focused on certain categories in any of the phases (e.g., \textit{Filtering Feedback} or \textit{Legal Reqs}). On the other hand, we also see that for some categories for which there is sufficient data, some interesting variations occur. In particular, most of the Original categories, namely \textit{Customer}, \textit{Facilities}, \textit{Workers} and \textit{Camp} tend to decrease. Conversely, the relevance of \textit{Communication} increases. In the other categories, interesting increments are observed for \textit{Search/Filter} and \textit{Usability}, while the relevance of \textit{Security/Privacy} and \textit{Advertisement} decreases. 

\begin{figure}[t]
\centering
\includegraphics[width=\columnwidth]{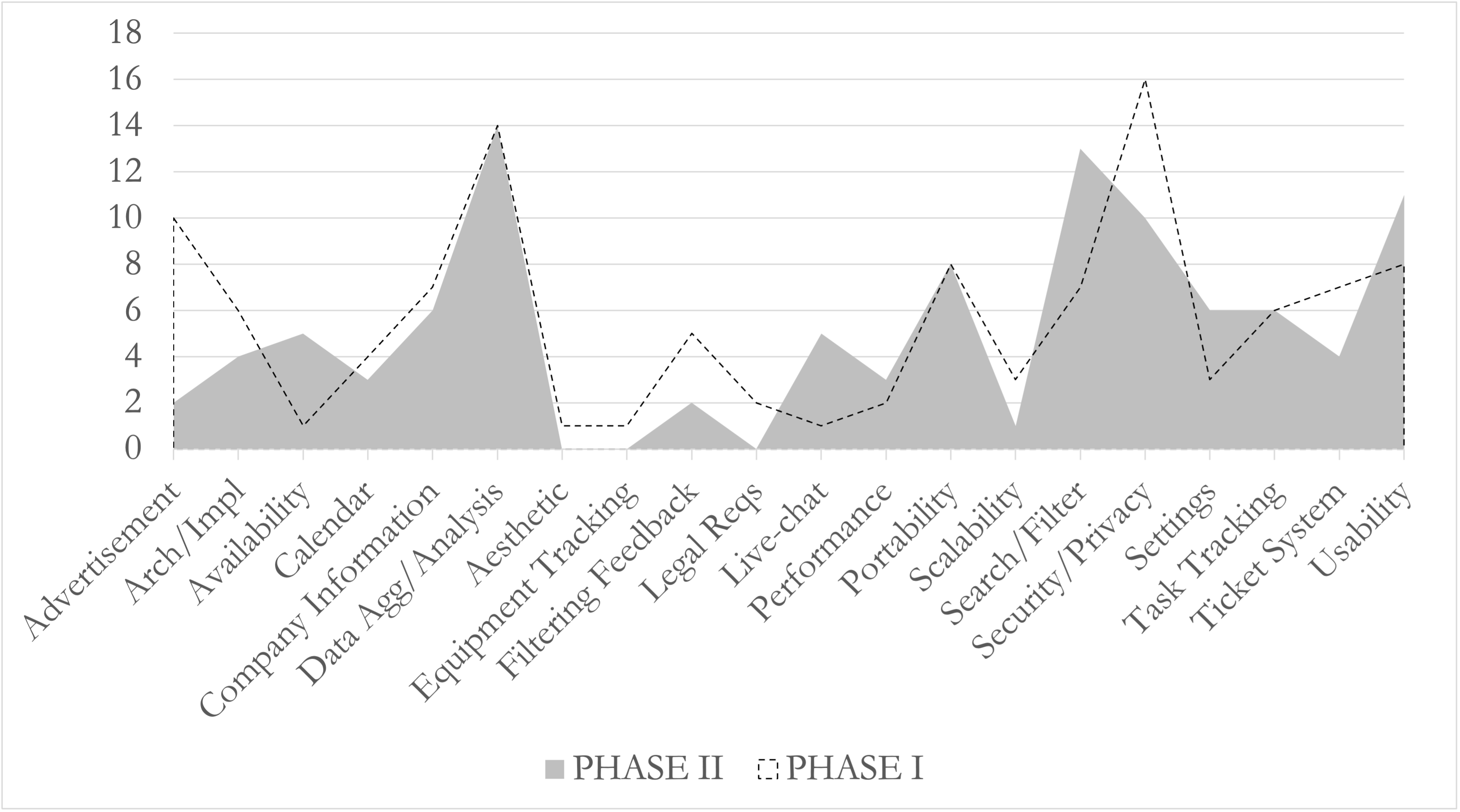}
\caption{Number of analysts who considered each interview category in Phase I and in Phase II.}
\label{fig:increasbyanalyst}
\end{figure}

Fig.~\ref{fig:increasbyanalyst} provides a different perspective, indicating the absolute number of analysts focusing on a certain category in Phase I and in Phase II. The diagram shows that some more analysts dedicated attention to features that were marginally considered before, such as \textit{Live-chat},  \textit{Settings}, and \textit{Availability}, while some topics are abandoned by part of the analysts in this phase, as, e.g., \textit{Ticket System} or \textit{Filtering Feedback}, either because they were considered completely addressed by the previously written requirements, or because these aspects did not appear as relevant in the retrieved apps.  

\begin{figure}[t]
\centering
\includegraphics[width=\columnwidth]{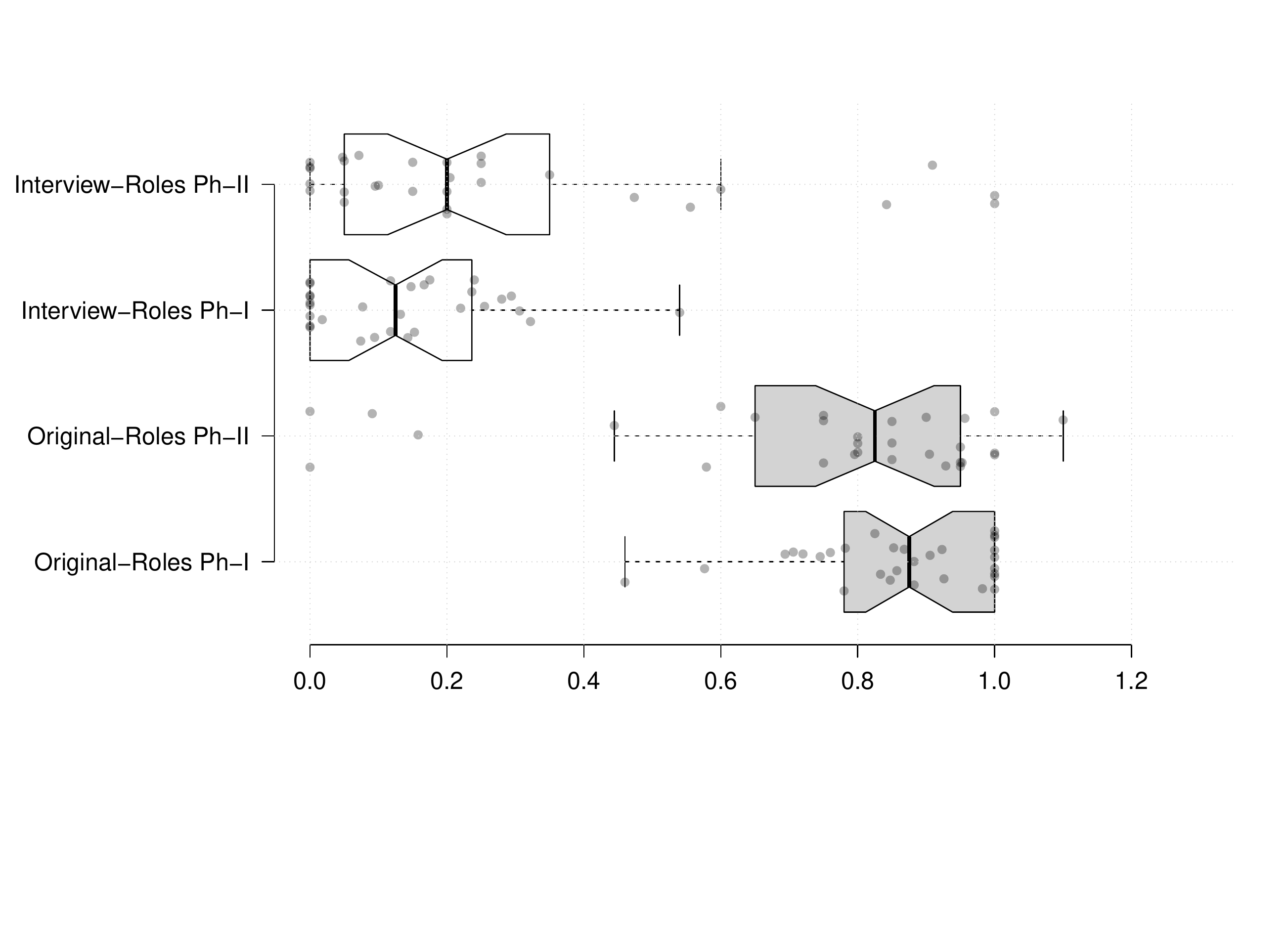}
\caption{\rev{Boxplots of the fraction of user stories after interviews (Ph-I) and after app store-inspired elicitation (Ph-II), considering different role groups, i.e., those in the original set (Original-Roles), and those identified after interviews (Interview-Roles). \textbf{Original-Roles-Ph-I/Original-Roles-Ph-II} = fraction of user stories using roles in the original set, in Ph-I and Ph-II, respectively; \textbf{Interview Roles-Ph-I/Interview-Roles-Ph-II} = fraction of user stories using roles introduced after interviews, in Ph-I and Ph-II, respectively. In a few cases, the fraction is higher than 1.0 because certain user stories were dedicated to more than one role. The figure indicates that in both Phase I and II, more user stories are dedicated to original roles, and fewer to new roles identified after interviews. Furthermore, in Phase II, we see that user stories dedicated to original roles decrease, in favour of those dedicated to new roles.}}
\label{fig:rolesboxplotvariation}
\end{figure}

\subsection{Analysis of Roles}

Concerning roles, Fig.~\ref{fig:rolesboxplotvariation} reports the difference between groups of roles in terms of rate of user stories dedicated to role groups in Phase I and Phase II. The grouping is analogous to the one already considered for user story categories. We do not report the boxplot for novel roles introduced after app store-inspired elicitation, since solely three subjects out of 30 introduced novel roles. From the figure, we see that the relevance of the original roles decreased, in favour of the roles identified after interview-based elicitation. The statistical tests reported in Table~\ref{tab:teststhreerolesgroups} show that the difference between the groups is significant for $\alpha = 0.5$ for \rev{Interview-CAT} roles. Therefore, there is a statistically significant variation in terms of relevance given to novel roles already identified in the interview phase. More specifically, interview roles increase by 100\% (the mean difference is 0.14, and the value in Phase I was 0.14). \rev{Instead, the mean decrease of 0.12 observed in terms of user stories dedicated to roles of the original set is not significant (p-value = 0.07).} 

\input{tables/tests-three-role-groups}

\begin{figure}[t]
\centering
\includegraphics[width=\columnwidth]{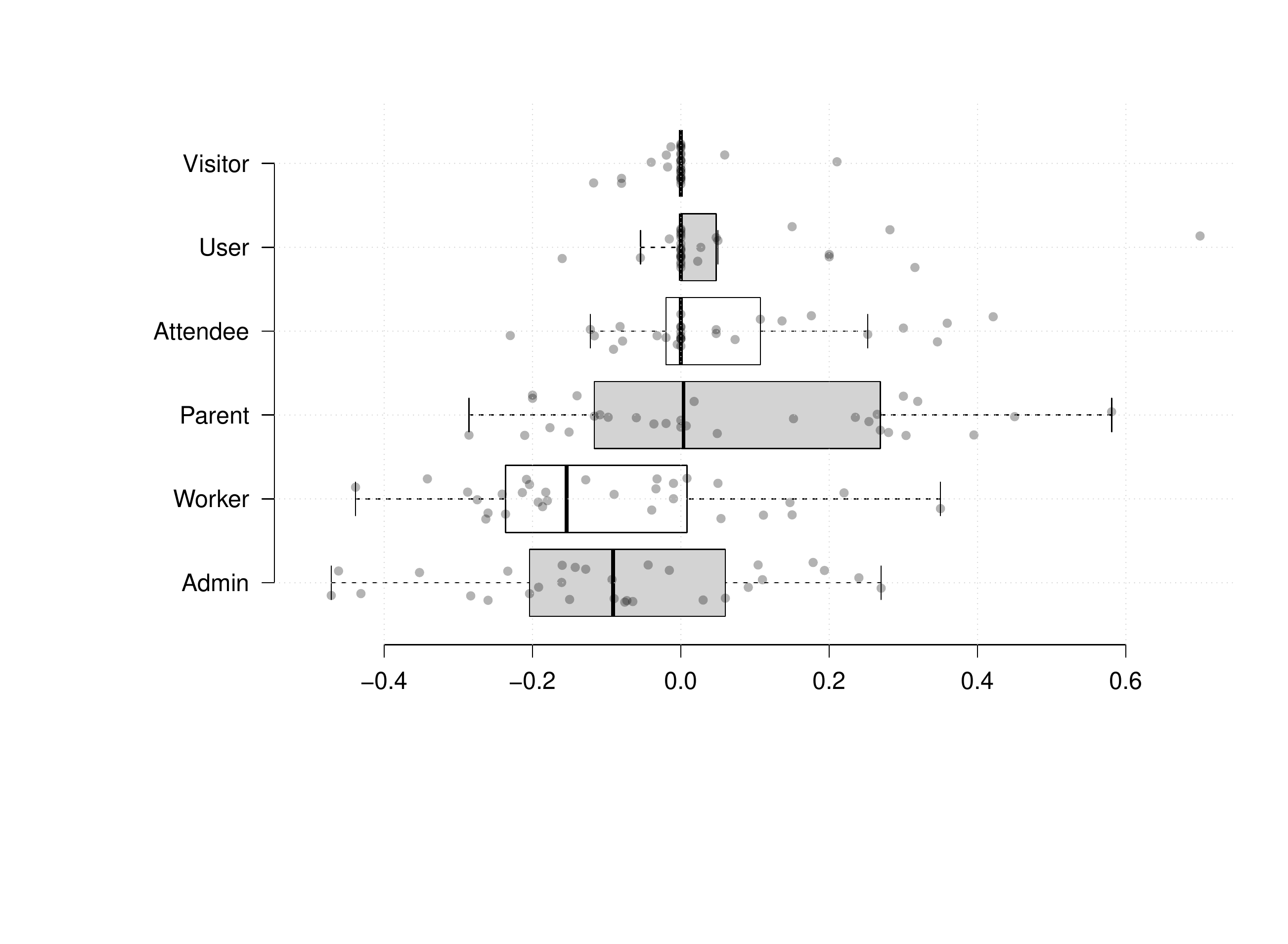}
\caption{Boxplots of the difference between the \rev{fraction} of user stories after app store-inspired elicitation (Phase II) and after interviews (Phase I), considering the different roles. The plot reports only the most common roles. \rev{The boxplots below 0 (\textit{Admin}, \textit{Worker}) indicate a decrease of relevance for that role in Phase II. The boxplots above 0 (\textit{Attendee}, \textit{User})} indicate an increase of relevance, as a greater fraction of user stories is dedicated to these roles.}
\label{fig:rolesboxplotbeforeafter}
\end{figure}

Concerning the relevance of specific roles, we can  qualitatively observe some variations in Fig.~\ref{fig:rolesboxplotbeforeafter}. The figure reports the distribution of the difference between Phase I and Phase II of the \rev{fraction} of user stories belonging to a certain role. We consider solely the most frequent roles. We see that the role of \textit{Admin} and \textit{Worker} decreased, while some increase is observed for the other roles. In particular, the generic role of \textit{User}, intended as app user, received more attention, together with \textit{Attendee} and \textit{Parent}. Overall, the app store analysis allowed the analysts to focus more on the user-side of the business, rather than on the internal view, \rev{represented by \textit{Admin} and \textit{Worker} roles}. 

\subsection{RQ2.3: What are the emerging categories and roles?}

\begin{figure}[t]
\centering
\includegraphics[width=\columnwidth]{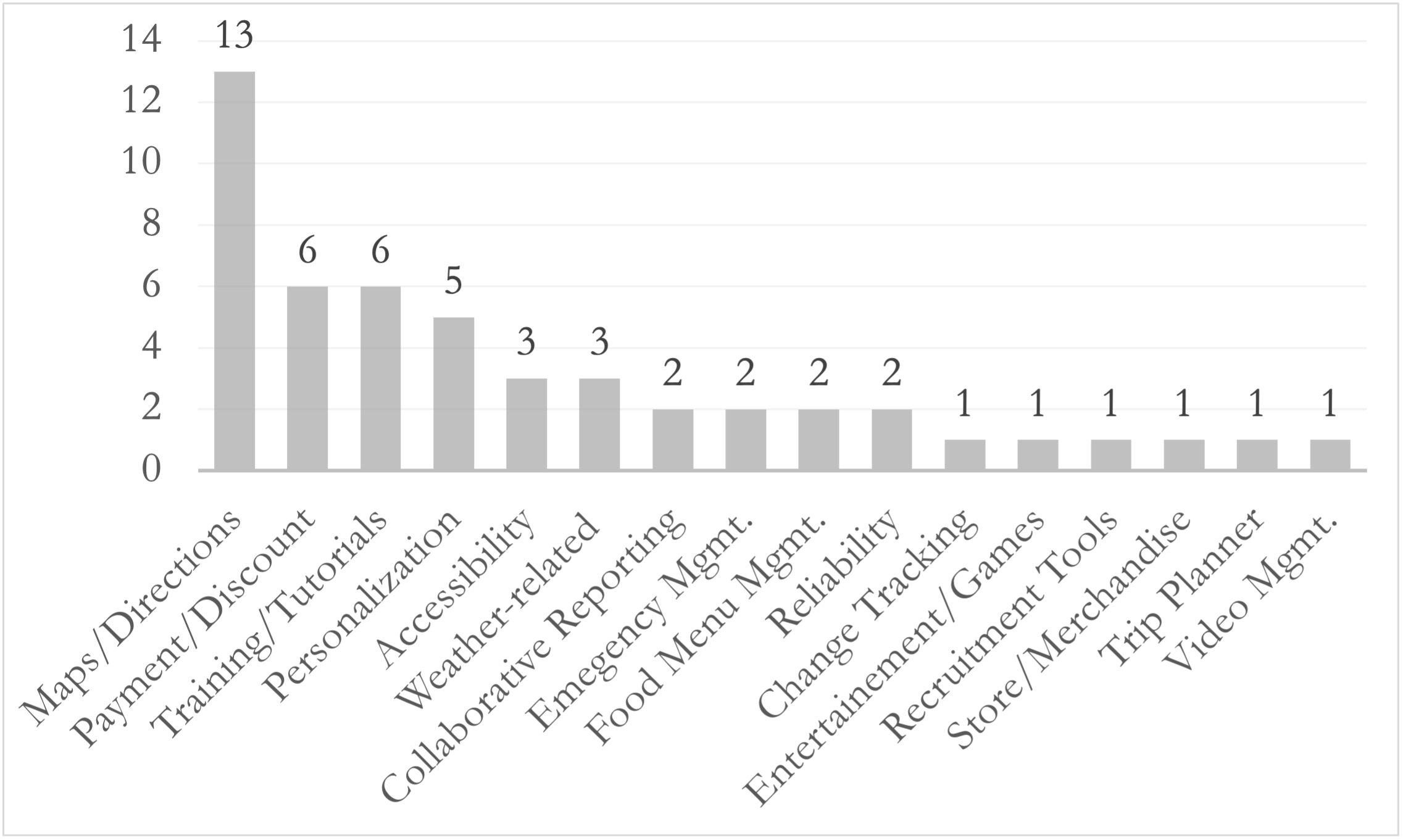}
\caption{Number of analysts using each novel category introduced in Phase II.}
\label{fig:categoriesphase2}
\end{figure}

 After app-store driven requirements elicitation, 80\% of the analysts introduced storied that did not belong to any of the previously identified categories and we have grouped these storied into 16 novel categories.
In Fig.~\ref{fig:categoriesphase2}, we report the histogram of these categories, considering the number of analysts that wrote at least one user story in the specific category. 

The most frequently used category, used by almost half of the analysts (13 out of 30), is \textit{Maps and Directions}, \rev{which contains all the stories related to the identification of positions on a map (e.g., for camp, event, individual), or to get directions}. An example is ``As a visitor, I want to be able to look for the event location using the application so that I do not have to drive through the campground looking for the best site available.''

Other frequently used new categories are \textit{Payment/Discounts} and \textit{Training/Tutorials}, both used by 20\% of the analysts.
They are both categories that introduce new functionalities in the system:
\rev{
\textit{Payment/Discounts} groups functionalities related to acquiring and using coupons, or understanding the costs of camps and activities (e.g., ``As a site user, I can add promotion code for the camp so that I can get discount on camp event''); \textit{Training/Tutorials} includes educational functionalities to train both the staff and the participants on the use of the app and on camp activities} (e.g., ``As a site user, I should see tutorial guide on how to book for app and how to use the system so it will help me to understand how to find camp events and make booking.'').

Finally, 5 analysts included stories to introduce the possibility to personalise the user experience in different ways. 
\rev{We have grouped these functionalities under \textit{Personalization}}. The descriptive statistics of these 4 more frequent categories are reported in Table~\ref{tab:NewCatPhaseII}.

\input{tables/descriptive-newCat-PhaseII}

Concerning roles, we have already observed that the role novelty rate is not negligible, meaning that individual analysts considered novel roles. However, most of these roles were already introduced by other analysts already in Phase I, and the number of entirely novel roles is limited to four (\textit{Outdoorsman}, \textit{Camping enthusiast}, \textit{Developer}, and \textit{Tourist}). Furthermore, only three analysts identified novel roles. Given these results, we cannot make relevant conclusions concerning the qualitative contribution of app store-driven elicitation towards the discovery of novel roles.

%% file: tables/tab-test-rates.tex
\begin{table}[]
\begin{tabular}{|l|l|l|l|l|l|}
\hline
\textbf{Rate}         & \textbf{\begin{tabular}[c]{@{}l@{}}Normality  \\      (Shapiro-Wilk)\end{tabular}} & \textbf{\begin{tabular}[c]{@{}l@{}}p-value \\      (Shapiro-Wilk)\end{tabular}} & \textbf{Method} & \textbf{L} & \textbf{H} \\ \hline
Conservation ($c$)      & Pass                                                                               & 0.40                                                                            & t-test          & 0.30       & 0.38       \\ \hline
Refinement ($r$)        & Pass                                                                               & 0.36                                                                            & t-test          & 0.54       & 0.62       \\ \hline
Novelty ($v$)           & Pass                                                                               & 0.08                                                                            & t-test          & 0.12       & 0.20       \\ \hline
Role Cons. ($c_{\rho}$) & Fail                                                                               & 0.03                                                                           & bootstrap       & 0.58       & 0.72       \\ \hline
Role Ref. ($r_{\rho}$)  & Fail                                                                               & 2.247e-06                                                                          & bootstrap       & 0.06       & 0.17       \\ \hline
Role Nov. ($v_{\rho}$)  & Fail                                                                               & 0.005                                                                          & bootstrap       & 0.18       & 0.29       \\ \hline
\end{tabular}
\caption{Results of the tests performed to identify the upper (H) and lower (L) bounds of the different rates.}
\label{tab:ratetest}
\end{table}

%% file: tables/descriptive-categories.tex
\begin{table}[]
\centering
\begin{tabular}{|l|c|c|c|c||c|}
\hline
\textbf{Category}      & \textbf{Mean} & \textbf{Std. Dev.} & \textbf{Min} & \textbf{Max} & \textbf{Original} \\ \hline
\textit{customer}      & 26.3\%       & 9.4\%             & 6.0\%       & 54.7\%      & 64.2\%           \\ \hline
\textit{facilities}    & 8.0\%        & 5.2\%             & 0.0\%       & 18.0\%      & 5.7\%            \\ \hline
\textit{personnel}     & 15.6\%       & 7.5\%             & 2.0\%       & 30.0\%      & 9.4\%            \\ \hline
\textit{camp}          & 9.9\%        & 8.0\%             & 0.0\%       & 30.1\%      & 7.6\%            \\ \hline
\textit{communication} & 32.0\%       & 10.6\%            & 9.6\%       & 48.2\%      & 15.1\%           \\ \hline
\end{tabular}
\caption{Descriptive statistics for the category distributions in the participants user stories, and the percentage in the original user stories.
}
\label{tab:categoryDistribution}
\end{table}

%% file: tables/test-initial-cats.tex
\begin{table}[]
\resizebox{1\linewidth}{!}{
\begin{tabular}{|l|l|l|l|l|l|l|}
\hline
\textbf{Category} & \textbf{Exp. $\mu$} & \textbf{Obs. $\mu$} & \textbf{Norm.} & \textbf{Test} & \textbf{Test Statistic} & \textbf{p-value}  \\ \hline
customer          & 64.15               & 26.34               & Pass           & t-test        & t = 15.263              & 2.14e-15          \\ \hline
facilities        & 5.66                & 8.04                & Fail           & wilcox        & V = 435                 & 2.69e-06          \\ \hline
personnel         & 9.43                & 15.60               & Pass           & t-test        & t = 11.467              & 2.71e-12          \\ \hline
camp              & 7.55                & 9.91                & Fail           & wilcox        & V = 406                 & 3.99e-06          \\ \hline
communication     & 15.09               & 31.95               & Pass           & t-test        & t = 16.586              & 2.44e-16 \\ \hline
\end{tabular}
}
\caption{Statistical tests to compare the mean of the distribution of the different categories (Obs. $\mu$) with respect with the one in the original set (Exp. $\mu$).}
\label{tab:testoriginalcats}
\end{table}

%% file: tables/descriptive-roles.tex
\begin{table}[]
\centering
\begin{tabular}{|l|c|c|c|c||c|}
\hline
\textbf{Role}          & \textbf{Mean} & \textbf{Std. Dev.} & \textbf{Min} & \textbf{Max} & \textbf{Original} \\ \hline
\textit{Administrator} & 41.9\%       & 14.8\%            & 11.3\%      & 83.3\%      & 66.0\%           \\ \hline
\textit{Worker}        & 27.4\%       & 12.3\%            & 0.0\%       & 62.8\%      & 94.3\%           \\ \hline
\textit{Parent}        & 16.9\%       & 9.2\%             & 0.0\%       & 32.0\%      & 24.5\%           \\ \hline
\textit{New Role}      & 13.8\%       & 13.4\%            & 0.0\%       & 54.0\%      & 0.0\%            \\ \hline
\end{tabular}
\caption{Descriptive statistics for the role distributions in the participants user stories, and the percentage in the original user stories.
}
\label{tab:roleDistribution}
\end{table}

%% file: tables/test-roles-phase-1.tex
\begin{table}[]
\resizebox{1\linewidth}{!}{
\begin{tabular}{|l|l|l|l|l|l|l|}
\hline
\textbf{Role} & \textbf{Exp. $\mu$} & \textbf{Obs. $\mu$} & \textbf{Norm.} & \textbf{Test} & \textbf{Test Statistic} & \textbf{p-value} \\ \hline
Administrator & 66.04               & 41.94               & Pass           & t-test        & t = 15.496                 & 1.45e-15        \\ \hline
Worker        & 9.43                & 27.38               & Pass           & t-test        & t = 12.226              & 5.75e-13         \\ \hline
Parent        & 24.53               & 16.92               & Pass           & t-test        & t = 10.075             & 5.57e-11         \\ \hline
\end{tabular}
}
\caption{Statistical tests to compare the percentage of the user stories in the different roles (Obs. $\mu$) with respect with the one in the original set (Exp. $\mu$).}
\label{tab:testrolesphase1}
\end{table}

%% file: tables/descriptive-newCat.tex
\begin{table}[]
\centering
\begin{tabular}{|l|c|c|c|c|}
\hline
\textbf{Category}            & \textbf{Mean} & \textbf{Std. Dev.} & \textbf{Min} & \textbf{Max} \\ \hline
\textit{Security/Privacy}    & 4.8\%        & 2.9\%             & 1.8\%       & 10.0\%      \\ \hline
\textit{Data Agg/Analysis}   & 3.9\%        & 2.1\%             & 1.9\%       & 7.8\%       \\ \hline
\textit{Advertisement}       & 2.7\%        & 1.1\%             & 1.7\%       & 4.0\%       \\ \hline
\textit{Portability}         & 4.2\%        & 2.0\%             & 2.0\%       & 8.3\%       \\ \hline
\textit{Usability}           & 6.9\%        & 2.7\%             & 2.0\%       & 20.0\%      \\ \hline
\textit{Company Information} & 4.0\%        & 3.6\%             & 1.8\%       & 11.9\%      \\ \hline
\textit{Search/Filter}       & 3.1\%        & 2.0\%             & 1.7\%       & 6.1\%       \\ \hline
\end{tabular}
\caption{Descriptive statistics summarizing the impact of the most recurrent emerging categories. \rev{The percentages are computed with respect to the total number of user stories of an analyst. \textbf{Mean}, \textbf{Std. Dev.}, \textbf{Min} and \textbf{Max} of these percentages consider only those analysts who actually wrote user stories belonging to the specific category.}}
\label{tab:NewCat}
\end{table}

%% file: tables/descriptive-NewRoles.tex
 \begin{table}[]
\centering
\begin{tabular}{|l|c|c|c|c|}
\hline
\textbf{Category}   & \textbf{Mean} & \textbf{Std. Dev.} & \textbf{Min} & \textbf{Max} \\ \hline
\textit{Attendee}   & 15.8\%       & 8.4\%             & 7.4\%       & 38.0\%      \\ \hline
\textit{Visitor}    & 7.1\%        & 4.0\%             & 1.8\%       & 11.9\%      \\ \hline
\textit{User}       & 14.2\%       & 12.0\%            & 3.4\%       & 35.9\%      \\ \hline
\textit{Consultant} & 4.7\%        & 3.1\%             & 2.0\%       & 8.0\%       \\ \hline
\end{tabular}
\caption{Descriptive statistics summarizing the impact of the most recurrent emerging roles. \rev{The percentages are computed with respect to the total number of user stories of an analyst. \textbf{Mean}, \textbf{Std. Dev.}, \textbf{Min} and \textbf{Max} of these percentages consider only those analysts who actually wrote user stories belonging to the specific role.}}
\label{tab:NewRole}
\end{table}


%% file: tables/tab-test-rates-apps.tex
\begin{table}[]
\begin{tabular}{|l|l|l|l|l|l|}
\hline
\textbf{Rate}          & \textbf{\begin{tabular}[c]{@{}l@{}}Normality  \\      (Shapiro-Wilk)\end{tabular}} & \textbf{\begin{tabular}[c]{@{}l@{}}p-value \\      (Shapiro-Wilk)\end{tabular}} & \textbf{Method} & \textbf{L} & \textbf{H} \\ \hline
Novelty ($v^{App}$)    & Pass                                                                               & 0.068                                                                        & t-test       & 0.28       & 0.42       \\ \hline
Refinement ($r^{App}$) & Fail                                                                               & 0.01739                                                                        & bootstrap       & 0.68       & 0.87       \\ \hline
Role Nov. ($v^{App}_\rho$) & Fail & 1.319e-05 & bootstrap & 0.08 & 0.17 \\ \hline
\end{tabular}
\caption{Results of the tests performed to identify the upper (H) and lower (L) bounds of the different app store-related rates.}
\label{tab:testratesapp}
\end{table}

%% file: tables/tests-three-groups.tex
\begin{table}[]
\resizebox{1\linewidth}{!}{
\begin{tabular}{|l|l|l|l|l|l|l|l|}
\hline
\textbf{Group}     & \textbf{Ph-I $\mu$} & \textbf{Ph-II $\mu$} & \textbf{Norm.} & \textbf{Test}   & \textbf{Difference} & \textbf{Test Stat.} & \textbf{p-value} \\ \hline
\textbf{Original-CAT}  & 0.92                & 0.71                 & Pass           & t-test (paired) & mean = -0.21         & t = -4.9181              & 3.185e-05         \\ \hline
\textbf{Interview-CAT} & 0.16                & 0.31                 & Pass           & t-test (paired) & mean = 0.14        & t = 4.5758             & 8.232e-05         \\ \hline
\textbf{App-CAT}       & 0                   & 0.14                 & Fail           & wilcox          & mean = 0.14         & V = 300                 & 1.88e-05         \\ \hline
\end{tabular}
}
\caption{Results of the statistical tests to verify whether the difference between the means of the distributions of the category groups in Phase I (\textbf{Ph-I} $\mu$) and Phase II (\textbf{Ph-II} $\mu$) is significant. \textbf{Norm.} = result of the normality test with Shapiro-Wilk; \textbf{Test} = applied test, based on the results of \textbf{Norm.}; \textbf{Difference} = mean difference; \textbf{Test Statistic} = value of the test statistic for the applied test.}
\label{tab:threegroupstest}
\end{table}

%% file: tables/tests-three-role-groups.tex
\begin{table}[]
\resizebox{1\linewidth}{!}{
\begin{tabular}{|l|l|l|l|l|l|l|l|}
\hline
\textbf{Group}     & \textbf{Ph-I $\mu$} & \textbf{Ph-II $\mu$} & \textbf{Norm.} & \textbf{Test}   & \textbf{Difference} & \textbf{Test Statistic} & \textbf{p-value} \\ \hline
\textbf{Original}  & 0.86                & 0.74                 & Fail           & wilcox (paired) & mean = -0.12         & V = 284                 & 0.07         \\ \hline
\textbf{Interview} & 0.14                & 0.28                 & Fail           & wilcox (paired) & mean = 0.14        & V = 111                  & 0.037         \\ \hline
\end{tabular}
}
\caption{Results of the statistical tests to verify whether there is a difference between the means of the distribution of the role groups in Phase I (\textbf{Ph-I} $\mu$) and Phase (\textbf{Ph-II} $\mu$). \textbf{Norm.} = result of the normality test with Shapiro-Wilk; \textbf{Test} = applied test, based on the results of \textbf{Norm.}; \textbf{Difference} = mean difference; \textbf{Test Statistic} = value of the test statistic for the applied test.}
\label{tab:teststhreerolesgroups}
\end{table}

%% file: tables/descriptive-newCat-PhaseII.tex
\begin{table}[]
\centering
\begin{tabular}{|l|c|c|c|c|}
\hline
                            & \textbf{Mean}    & \textbf{Std. Dev.} & \textbf{Min}    & \textbf{Max}     \\ \hline
\textit{Maps/Directions}    & 11.71\% & 7.75\%    & 4.00\% & 31.82\% \\ \hline
\textit{Payment/Discount}   & 6.99\%  & 3.68\%    & 4.55\% & 31.82\% \\ \hline
\textit{Personalization}    & 8.98\%  & 3.54\%    & 5.26\% & 21.05\% \\ \hline
\textit{Training/Tutorials} & 5.63\%  & 1.39\%    & 4.55\% & 21.05\% \\ \hline
\end{tabular}
\caption{Descriptive statistics summarizing the impact of the most recurrent novel categories emerging from Phase II.  \rev{\textbf{Mean}, \textbf{Std. Dev.}, \textbf{Min} and \textbf{Max} consider the percent rate of user stories in a certain category with respect to the total number of user stories produced in Phase II. \textbf{Mean}, \textbf{Std. Dev.}, \textbf{Min} and \textbf{Max} of these percentages consider only those analysts who actually wrote user stories belonging to the specific category.}}
\label{tab:NewCatPhaseII}
\end{table}

%% file: sections/discussion.tex
\section{Discussion}
\label{sect:discussion}
The analysis of the data collected in our study suggests the presence of interesting patterns in the analysts' behavior that empirically confirm the intuition that requirements are not only elicited but co-created \rev{through the interview process. Then, requirements are substantially extended when the analyst looks at similar products in the market. Overall, our analysis shows that throughout the elicitation process there is an \textit{evolution} of the original idea.} While this evolution might be partially driven by the three  Cs~\cite{ZOWGHI2003interplay}, our data show that it does not only go in the direction of completing the existing information, but often changes the focus of the requirements and the roles, adds new functionalities that were not part of the initial ideas, and introduces nonfunctional requirements. In the following, we answer the RQs and we provide observations in relation to existing literature. \rev{Our study contributes to \textit{theory} in RE, and it is mainly oriented to researchers. In the following, we use the $\Rightarrow$ symbol to highlight take-away messages to trigger future studies of the RE community.}

\vspace{10pt}
\begin{mdframed}[style=style1]
\textbf{RQ1.1} There is a substantial difference between initial customer's ideas and the requirements documented after requirements elicitation interview sessions. Up to 63\% of the documented requirements are dedicated to refinement of initial ideas, and up to 20\% to entirely novel features. Involved roles are generally preserved, but up to 29\% of the requirements cover novel roles.
\end{mdframed}
\vspace{10pt}

\textit{Observations.} These data quantitatively show the paramount contribution of the interview-based elicitation to the final content of the requirements document. In particular, this process mostly focuses on \textit{refining} initial ideas. 
Existing literature have provided theoretical frameworks towards understanding requirements elicitation~\cite{ZOWGHI2003interplay,gervasi2013unpacking}. Other studies have focused on interviews~\cite{Davis2006}, showing the role of domain knowledge~\cite{Hadar2014Role,NiknafsB17}, communication skills~\cite{coughlan2002effective}, and possible mistakes~\cite{bano2019teaching}. 
This is the first study that quantitatively provides evidence on the role of interview-based requirements elicitation, and its impact on documented requirements. It shows that evolution of requirements starts right after their expression, and that a relevant number of features that were not initially conceived are introduced, even after two short interviews. This evolution has been enabled by the interview process, as dialogue is a primary trigger to establish a common understanding~\cite{Sutcliffe2013}, and also facilitates creativity. On the other hand, we argue that a crucial role was played by the usage of diagrams and mock-ups, which the analysts could use to support early analysis and to show ideas to the customer, as suggested by common practice~\cite{Sutcliffe2013}. Finally, the background, preliminary research, and creativity of the analyst can have contributed to this result. From our study, we do not know which of these different factors contributed the most to the final content of the requirements.

$\Rightarrow$        Future research should focus on measuring what are the actual factors that impacted the most on early requirements evolution, whether dialogue, diagrams and prototypes, background, or creativity potential of the involved subjects.

\vspace{10pt}
\begin{mdframed}[]
\textbf{RQ1.2} The relevance given to specific requirements categories substantially shifts between initial customer's ideas and documented requirements. Less relevance is given to customer-related requirements, with a decrease of about 59\%, in favour of other categories and roles.   
\end{mdframed}
\vspace{10pt}

\textit{Observations.} 
In our study, there were three main roles associated to requirements, which were in general maintained in the analysts' user stories. 
However, while in the original specification (and thus in the mind of the customer) the dominant perspective was the one of the administrator---i.e., the fictional customer---, the majority of the analysts re-balanced the focus among different perspectives and introduced also additional ones. This is very interesting especially because the analysts only spoke with the administrator. This suggests that stakeholder  identification, a crucial phase of requirements elicitation~\cite{PACHECO2012systematic}, is second nature to the analysts. This, however, can also create issues when certain potential system users are not consulted. Identifying the relevant users is not sufficient, as these need also to be interviewed to make sure to correctly represent their perspective.  

At this regard, it is curious to observe that the majority of the analysts considered the perspective of the camps' participants and imagined them as users of the system. Camps are usually available for a large range of ages including young children who most likely will not own a mobile device and so will not be a direct user of the system. 
When analysts are not knowledgeable of the domain or do not prepare enough~\cite{bano2019teaching}, they might include requirements that are inappropriate for the context.


This adds to the conclusion of Hadar \textit{et al.}~\cite{Hadar2014Role}, who observed that domain knowledge helps to direct the requirements elicitation process, but also makes it difficult to listen to customers. Here we see that not only some customer's needs may be neglected, but some additional requirements may be introduced that are not appropriate for the context. This observation highlights the importance of validating requirements with the system's stakeholders to gather their feedback also during elicitation and the early stages of modeling and specification~\cite{Leite91Req}. It also highlights that systematic processes need to be established for validation, as unstructured meetings are not sufficient to ensure that requirements are correctly collected, i.e., all requirements are collected, and no unwanted requirements are introduced.

$\Rightarrow$           Analysts might include requirements derived from their domain experience, and exclude relevant ones. Systematic \textit{validation} of the requirements need to be enforced to ensure that all relevant requirements are collected, and novel requirements are agreed upon. \rev{Future research should focus on estimating the impact of requirements validation on the final requirements document.}

\vspace{10pt}
\begin{mdframed}[]
\textbf{RQ1.3} Several novel categories of requirements emerge during the elicitation process. Requirements related to privacy and security, as well as other non-functional aspects, such as advertisement, portability and usability are common to a relevant number of analysts \rev{(i.e., from 8 to 16, out of 30 analysts)}. Dominant functional aspects are related to data aggregation, information access, and introduction of features inspired by the software engineering process.
\end{mdframed}
\vspace{10pt}

\textit{Observations.} A considerable part of the new requirements introduced by the analysts are nonfunctional requirements (e.g., privacy, usability, portability, scalability) and this could be a sign of the analysts' maturity \rev{and competence---also acquired through their university courses}. We argue that analysts' training and expertise could drive them to ask questions related to nonfunctional aspects, which could be initially overlooked by the customer, but that are recognised to be vital to the success of the product~\cite{glinz2007non,chung2012non}. Indeed, as shown by Pitts and Browne~\cite{Pitts2004Stopping}, expert analysts use their belief structure for the construction of mental lists of topics they want to cover in the elicitation process and, among them, are nonfunctional requirements that require expertise and so are often not mentioned by the customers. 
Therefore, we can conclude that analysts contribute with the elicitation and documentation of nonfunctional requirements, which are not considered by the customer.

Analysts also introduce functional features that come from their background as computer scientists. Ticket systems, search and filtering features, setting options, and support for data aggregation and analysis, are typically used by software engineers and system administrators. These are somehow ``ported'' by analysts from the software engineering domain to the specific application domain. In other terms, the analysts look at the product as it was going to be used by someone who has a software engineering mindset. The design of systems according to the viewpoint of computer scientists, with the consequent shaping of the world in terms of software engineering practices and principles, has been theorised by Baricco in his essay ``The Game''~\cite{baricco2019game}. In our experiment, we observe this theory in action, and we are not aware of other studies in RE in which this aspect emerged. 

$\Rightarrow$           Further research is needed to answer the question: \textit{To what extent software engineering practices and principles shape the functionalities supported by software products?} 

\vspace{10pt}
\begin{mdframed}[]
\textbf{RQ2.1} There is a substantial difference between the requirements documented after interview sessions, and those documented after app store-inspired elicitation. Up to 42\% of the requirements belong to categories that were not previously considered by the analyst, and up to 17\% novel roles can be introduced. 
\end{mdframed}
\vspace{10pt}

\textit{Observations.} App-store inspired elicitation clearly leads to additional requirements evolution, and allows analysts to discover novel features that were not considered in the previous phase. Its contribution to the final set of requirements is therefore strongly relevant, and the \rev{analysis of similar products in the market} should be a mandatory step in requirements elicitation. App store-inspired elicitation is frequently practiced by app developers, to check what is the position of their product in the market, and also to feed the process of idea generation during the product concept phase~\cite{al2019app}. While previous work focused on supporting this activity with automated recommendation  tools (see, e.g.,~\cite{chen2019recommending,liu2019information,jiang2019recommending}), this is the first work that quantitatively shows how much is the additional contribution to documented requirements given by browsing similar products. \rev{Furthermore, the recent survey by Dabrowski \textit{et al.}~\cite{dkabrowski2022analysing} has observed that limited evidence is available to show that mining the app store is actually useful for software engineering. Our work provides this missing evidence, also showing \textit{how much} this can be useful.} \rev{In the many development contexts in which app stores are not available (e.g., enterprise software, business, embedded software), different forms of market analysis should be practiced, to get inspiration from similar products. Studies on natural language processing (NLP) tools to support these tasks have been performed---e.g., based on the analysis of brochures of competing products~\cite{ferrari2013mining} or online descriptions~\cite{davril2013feature,dumitru2011demand}---and could provide a valid alternative to support market analysis in these contexts.}

In our evaluation, we did not specifically analyse what are the app store browsing and selection strategies that contributed the most to the observed requirements evolution. Future research should investigate these aspects, to devise practical strategies of analysts, which can in turn inform the development of recommender systems.

$\Rightarrow$ It should be also noted that, in this study, app store-inspired elicitation comes after interviews, and we do not know what could be the results if the two phases were swapped, or if just app store-inspired elicitation was carried out. Future studies are needed with a within-subjects crossover design (to compare order effects), and with a between-subjects design (to compare the two elicitation styles in parallel), to clarify these aspects.

\vspace{10pt}
\begin{mdframed}[]
\textbf{RQ2.2} The relevance given to specific requirements categories shifts between previously documented requirements and additional ones introduced after app store-inspired elicitation. The categories that were part of the initial ideas still dominate on average (71\% of requirements), but become less relevant by 23\%. More relevance (increase by 45\%) is given to categories that were introduced after interview-based elicitation, and in particular to usability aspects. A similar behaviour is observed also for roles, with an increase of requirements dedicated to roles introduced after interview-based elicitation (100\% increase), and in particular the generic ``user'', and a decrease of attention towards customer-related roles (20\%). 
\end{mdframed}
\vspace{10pt}

\textit{Observations.} 
The app store-inspired elicitation activity helps analysts to refine features that were developed during interview-based elicitation, and leads to further evolution of the requirements. The visualisation of realised products, and the interaction with them---we recall that analysts were asked to download and try the products---helps to better focus on usability aspects, and to understand what could be the possible usability needs of future users. App store-inspired elicitation helps to put more emphasis on the generic role of ``user'', rather than on the needs of the customer, thereby expanding also the potential public of the product itself. Furthermore, requirements concerning \textit{communication}, already well elaborated in the previous phases, become even more important now. Many apps are actually focused on communication, information sharing and social media-related aspects and this could have triggered higher attention on  this topic. It is also interesting to notice that security and privacy features, introduced after interview-based elicitation, becomes less relevant in this phase. This suggests that these types of requirements are not triggered by the app browsing activity, despite their relevance for app development recognised by the abundance of research in the area, e.g., concerning privacy policies~\cite{slavin2016toward,hatamian2020engineering}. 

Apparently, the introduction of security and privacy requirements is strictly a contribution of the analysts' competence, as it was not initially triggered by the customer, and it is not emphasised further by app store-inspired elicitation. 

It is also interesting to contrast the answer to RQ2.2 with the results of RQ2.1. Indeed, RQ2.1 focuses on the novelty rate for specific analysts---i.e., a user story was considered as \textbf{A} if its category was novel for the analyst. However, the category could also be identified previously by other analysts. Instead, RQ2.2 considers aggregate data, i.e., the contribution of the specific analyst to the whole set of novel categories. RQ2.1 shows that the novelty rate is surprisingly high, while RQ2.2 shows that the greater increase concerns categories that were already identified by some analysts in the previous phase. This means that categories are discovered during app store-inspired elicitation, but a substantial part of them were already discovered by other analysts through interviews.

$\Rightarrow$ Therefore, interview-based elicitation or app store-inspired elicitation frequently lead to categories in the same set. One elicitation approach or the other can be more appropriate, depending on the analyst's attitude, to discover certain requirements categories. This reinforces the need to apply both the approaches to produce a complete requirements document.

\vspace{10pt}
\begin{mdframed}[]
\textbf{RQ2.3} Novel categories of requirements are produced after app store-inspired elicitation. In particular, the review of apps enabled the introduction of functional features not considered before, and related to maps and direction, payment options, and personalisation.
\end{mdframed}
\vspace{10pt}

\textit{Observations.} Besides communication-related and social media aspects, one of the crucial features of mobile apps is geo-localisation. Aspects related to maps and directions are therefore naturally triggered by browsing apps. As noticed by one of the respondents of the survey of Al-Subaihin \textit{et al.}~\cite{al2019app}, users have been accustomed to a set of features that become \textit{de facto} a must for a new project, and geo-localisation related features are among these ones. 

Browsing the market also increases attention to market-related issues, and in particular payment options. Looking at the different payment plans made available by apps, and considering the feedback of users at this regard, can highly help the analysts to design an appropriate business model for their product. 

$\Rightarrow$ The presence of payment-related aspects in the novel requirements is particularly interesting, as business models of mobile apps are still an emerging research area~\cite{cristofaro2020business}. Out results show that browsing similar apps does not only help to validate initial ideas, but can also help to plan for a profitable  business.

%% file: sections/threats.tex
\section{Threats to Validity}
\label{sect:threats}
\paragraph{Construct Validity} The main constructs of interest are ``initial ideas'', and ``documented requirements''---after interview-based, and after app store-inspired elicitation. The construct of ``initial ideas'' is a somewhat vague concept, which is strictly related to the notion of \textit{pre-requirements} introduced by Hayes \textit{et al.}~\cite{hayes2008prereqir}. We \textit{reify} the intuitive meaning of this concept through a form that is well defined in the literature, and that can make it comparable with the construct of ``documented requirements'', namely \textit{user stories}. In analysing initial ideas---and their counterpart, documented requirements---through the user story representation, we consider multiple rate variables that are related to subjective evaluations. We mitigate subjectivity threats through the triangulation process described in Sect.~\ref{sec:datacollection}. 

The list of user stories that was used to reify initial ideas was written beforehand by other authors, and not by our fictional customer. In other terms, what we actually use is his understanding of the ideas of someone else, which is not the construct we are interested in. However, even in real contexts, the subject who speaks with a requirement analyst is often someone who has collected different ideas from other subjects. Furthermore, if we would ask our customer to write down user stories for his own initial ideas, the act of writing would be a further bias, as the documented ideas would not be ``initial'' anymore.  Given these limitations, also due to the complexity of the considered problem, we argue that our design represents an acceptable trade-off between construct validity and potential bias that could be introduced with a different design. 

In relation to construct validity, it should be noticed that we consider the contribution of app store-driven elicitation only in the form of \textit{extension} of the documented requirements. Indeed, the analysts could not modify the user stories they previously wrote, but only add further ones. We made this choice to have a more manageable design. Our conclusions about the contribution of app store-driven elicitation therefore do not apply to edit or deletion operations that could realistically occur. 

We also notice that the extension of requirements after app store analysis is performed \textit{before} the software is developed. This is a realistic assumption, as the survey of  Al-Subaihin \textit{et al.}~\cite{al2019app} that 65\% of developers browse the app stores with the purpose of validating the app's idea.

\rev{A final threat to construct validity concerns the indication  given to analysts on the number of user stories to be produced in the two phases (50 to 60, and 20, respectively). This was due to the need to obtain comparable samples---also considering the number of original user stories---so as to facilitate statistical analysis. It should be noted that the numbers were given as an indication, and not as a strict upper or lower thresholds, and analysts actually produced user stories in the neighborhood of the indicated values. We acknowledge that some analysts could have added user stories to reach the suggested numbers. Nevertheless, even in these cases, we argue that the observed rates and categories are still representative of the \textit{main point of attention} of the analysts during requirements documentation---i.e., original ideas, novel ideas, or app-store inspired ideas---which is what our study wants to investigate.} 

\paragraph{Internal Validity}

The initial user stories were studied by the fictional customer, and represent his interpretation of these initial ideas, which cannot be considered entirely faithful. Furthermore, given the repetition of interviews involving the same customer, a learning bias could not be entirely avoided. These elements could lead to a partial discrepancy between the user stories and what the customer communicated to the analysts during the interviews. However, the fictional customer is a trained research assistant and was asked to provide uniform interviews to the different analysts. He was asked to stick as much as possible to the initial user stories, while limiting further elaboration to the questions asked by analysts, as it would happen in a real setting. We believe that this is a sufficient countermeasure in the adopted context to guarantee uniform treatments to all analysts. 

The choice of the customer as one of the two researchers who performed the analysis of the user stories might create bias, as well as asymmetry between the results of the two researchers performing the analysis. However, the validity procedure for the analysis includes many reconciliation moments which had the goal of mitigating these threats.

One additional threat affects the app store-inspired phase. Specifically, it is not possible to ensure that the user stories produced after this second phase are entirely due to the app store searches. Indeed, the analysts could have included content that was derived from additional reflections on previous activities, or could have extended the requirements simply based on their intuition and experience. To mitigate this threat, we asked analysts to explicitly report the selected similar apps from which they took inspiration, and to explain why these were relevant, and what features could be adapted to their context. We believe that this exercise enabled them to better focus on the features of similar apps, therefore making them the dominant driver for the extension of the requirements.

\paragraph{External Validity}
Given that this research is a \textit{laboratory experiment}, intended as a study oriented to \textit{identify the relationship between several variables or alternatives under examination}~\cite{stol2018abc}, external validity is inherently limited. A limited number of user stories was used compared to a real system, and a single system was considered which is not necessarily representative of all types of systems. \rev{In particular, our experimental setting is representative of systems that can have a mobile version, or include a relevant mobile app-oriented component, so that comparison with
products in the app store market can be possible. Other systems---e.g., enterprise, business, embedded, etc.---do not have an app store, and comparison with similar, or competing products need to use different strategies~\cite{ferrari2013mining,dumitru2011demand}.}
\rev{We also acknowledge that different results could be obtained with another type of system, even if this  has a mobile app component. However, our quantitative results should be read as indicators of a trend, and not interpreted as absolute values. Future families of experiments, possibly including different app-oriented systems, will be carried out to consolidate our findings.}

\rev{Given the constrained nature of experiments}, different results may be obtained in a more realistic setting. \rev{However, in a realistic setting some variables could hardly be controlled (e.g., type of system, amount of user stories produced), especially if we wish to guarantee uniform treatments, and statistical significance}. \rev{Differently from a realistic setting, our study did not include a requirements validation step, in which elicited requirements are confirmed or rejected by the stakeholders. This was driven by the need to focus on the impact of analysts' contributions to the requirements. In future studies, we plan to focus on the impact of requirements validation, so that a more complete picture can be provided.} 
We used students instead of professionals in our experiment, as it is common and widely accepted in software engineering research~\cite{svahnberg2008using,falessi2018empirical,salman2015students}. Most of the involved participants also work as professional developers or analysts. 

The length of the interview might be limited with respect to the size of the system. However, this choice is in line with similar studies having a comparable setting (e.g., \cite{ferrari2020sapeer, bano2019teaching}) and, in addition, we performed two 15 minutes interviews, that given the preparation of the interviewee and the notwithstanding the limitations of the experimental context, can be considered sufficient to deliver complete information about the chosen system.

The use of a single stakeholder to present the point of view of different roles also represents a threat for our study and we acknowledge that different results could have been obtained if multiple stakeholders were interviewed.

\rev{
\paragraph{Conclusion Validity} Conclusion validity concerns the statistical analysis of results. We verified the assumptions of normality before applying parametric tests, and numerical evidence has been provided. Non-parametric tests were used when the assumptions could not be quantitatively verified. A residual threat concerns the multiple comparison problem~\cite{abdi2007bonferroni}. Indeed, in our study we perform multiple related tests on the same data, and we acknowledge that some observed differences could be due to chance. However, this possibility is limited for those cases in which the \textit{p-value} is particularly low, i.e., Table~\ref{tab:testoriginalcats},~\ref{tab:testrolesphase1}, and ~\ref{tab:threegroupstest}, while it is present for Table~\ref{tab:teststhreerolesgroups}---a Bonferroni correction would lead to $\alpha = 0.025$, which would lead to inconclusive results for both rows of the table.
}



%% file: sections/conclusion.tex
\section{Conclusion}
\label{sect:conclusion}

Requirements start from unexpressed ideas to be transformed into documented needs, and eventually realised into (\textit{satisfied by}) specifications and products. 
Understanding the evolution of requirements at their early stages can contribute to understand what are the most appropriate elicitation strategies to adopt. In this paper, we study early requirements evolution, when they pass from initial customer ideas into documented needs, and then are extended through the analysis of similar products from the app stores. To this end, we perform a laboratory experiment involving 30 subjects, and we quantitatively and qualitatively evaluate this evolution. Our study shows that the elicitation and documentation process can be regarded as a \textit{co-creation activity} involving the contribution of analysts and customers alike. The analyst's creativity becomes then central when analysing similar products to take inspiration from them.
The process does not only complete the initial ideas, but often changes the relevance given to specific requirements and roles, adds new functionalities, and introduces nonfunctional requirements. 
\rev{Our work contributes to \textit{theory} in RE}, and should be regarded as an empirically grounded starting point to better understand the transition from ideas into products. Furthermore, our work is also the first one in which traditional interview-based elicitation and app store-inspired elicitation are combined, showing how these provide complementary contributions to the final content of the requirements document. \rev{Finally, our work is also the first one that provides quantitative evidence of \textit{how much} app store mining can be useful for RE, thus addressing part of the limitations observed by Dabrowski \textit{et al.}~\cite{dkabrowski2022analysing} concerning the limited evidence of practical utility of this practice.} 

\rev{It is worth remarking that an experiment as inherent limitations~\cite{stol2018abc} that hamper generalisation of the results. While the specific quantitative results observed may be different in real contexts, the main messages of our study remain valid: \textit{1. in interview-based elicitation, requirements are not elicited but co-created by stakeholders and analysts}; \textit{2. interviews and app store analysis play complementary roles in requirements definition}; \textit{3. app store analysis provides a relevant contribution in practice}.}

At this stage, we looked into the elicitation and documentation process as a black box, and did not investigate the impact of the different means (mock-ups, prototypes) used for the analysis, and their relationship with the results. Future work oriented to unpack this black-box will address this issue. \rev{Furthermore, our work focused on quantitative analysis, and did not investigate the types of apps selected by the analysts, the strategies adopted to identify them, or the viewpoints of participants of advantages and disadvantages of the different elicitation techniques. These aspects will be studied in future work.  
Finally}, we also aim to observe the further evolution of the requirements into the actual products developed, to have a complete trace of their transformation. This study can serve as a baseline to support future automated software engineering methods oriented to manage requirements evolution. 

